\documentclass[journal]{IEEEtran}
\usepackage{graphicx}
\usepackage{cite}
\usepackage{picinpar}
\usepackage{amsmath}
\usepackage{url}
\usepackage{flushend}
\usepackage[latin1]{inputenc}
\usepackage{colortbl}
\usepackage{soul}
\usepackage{multirow}
\usepackage{pifont}
\usepackage{color}
\usepackage{alltt}
\usepackage[hidelinks]{hyperref}
\usepackage{enumerate}
\usepackage{siunitx}
\usepackage{breakurl}
\usepackage{epstopdf}
\usepackage{pbox}
\usepackage{braket}
\usepackage{stfloats}
\usepackage{amsmath}
\usepackage{amssymb}
\usepackage{blindtext}
\usepackage{subfig}
\usepackage{tabu}
\usepackage{multirow}
\usepackage{comment}

\usepackage{amsmath}
\usepackage{amsthm}
\usepackage{amssymb}
\usepackage{mathrsfs,bm}
\usepackage{graphicx}
\usepackage[linesnumbered,ruled,vlined]{algorithm2e}

\usepackage{booktabs}
\usepackage{multirow}
\usepackage{siunitx}
\usepackage{algorithmicx}

\usepackage{algpseudocode} 
\usepackage{multirow}
\usepackage{bbm}
\usepackage{arydshln}
\usepackage{wrapfig}
\usepackage[normalem]{ulem}
\usepackage{tabu}
\usepackage{multicol}
\usepackage{tikz}
\usetikzlibrary{arrows.meta, positioning}
\usepackage{makecell}
\usepackage{tikz}
\newcommand{\ignore}[1]{}

\newcommand{\classifier}{\mathtt{C}}

\newcommand{\match}{\mathtt{MATCH}}
\newcommand{\mismatch}{\mathtt{MISMATCH}}
\newcommand{\scores}{\mathbf{s}}

\newcommand{\name}{\textcolor{black}{SQuaD~}}

\usepackage{amsmath, amssymb, amsthm}
\usepackage[normalem]{ulem}

\begin{document}
\title{\name : Smart Quantum Detection for Photon Recognition and Dark Count Elimination}

\author{
	\vskip 1em
	
	Karl C. Linne (Kai Li), Sho Uemura, Yue Ji, 
	 Allen Zang,  Martin Di Federico,  Orlando Quaranta, Gustavo Cancelo, Debashri Roy
 \\

	\thanks{
	
		
		 Karl C. Linne (Kai Li),  and Allen Zang are with Pritzker School of Molecular Engineering,  University of Chicago, Chicago, 60637 , USA. (e-mail: karlchrislinne@gmail.com). 
         \\

         Sho Uemura, Martin Di Federico, and Gustavo Cancelo are with Fermi National Accelerator Laboratory, Batavia, 60510, USA. .
         \\

         Yue Ji with New York University, New York, 10012; USA. Orlando Quaranto is with Argonne National Laboratory, Lemont, 60439.
		
	}
}

\maketitle
	
\begin{abstract} 
Quantum detectors of single photons are an essential component for quantum information processing across computing, communication and networking.   
Today's quantum detection system, which consists of single photon detectors, timing electronics, control and data processing software, is primarily used for counting the number of single photon detection events. However, it is largely incapable of extracting other rich physical characteristics of the detected photons, such as their wavelengths, polarization states, photon numbers, or temporal waveforms.
This work, for the first time,  demonstrates a smart quantum detection system, \name, which integrates a field programmable gate array (FPGA) with a neural network model, and is designed to recognize the features of photons and to eliminate detector dark-count. The \name is a fully integrated quantum system with high timing-resolution data acquisition, onboard multi-scale data analysis, intelligent feature recognition and extraction, and feedback-driven system control. Our \name
experimentally demonstrates
1) reliable photon counting on par with the state-of-the art commercial systems; 
2) high-throughput data processing for each individual detection events;
3) efficient dark count recognition and elimination; 4) up to 100\% accurate feature recognition of photon wavelength and polarization. 
Additionally, we deploy the \name to an atomic (erbium ion) photon emitter source to realize noise-free control and readout of a spin qubit in the telecom band, enabling critical advances in quantum networks and distributed quantum information processing.
\end{abstract}


{}

\definecolor{limegreen}{rgb}{0.2, 0.8, 0.2}
\definecolor{forestgreen}{rgb}{0.13, 0.55, 0.13}
\definecolor{greenhtml}{rgb}{0.0, 0.5, 0.0}

\section{Introduction}
\label{sec:intro}
\IEEEPARstart{Q}{uantum} networks~\cite{kimble2008quantum,wehner2018quantum} hold the promises for long-distance secure communication\cite{li2023bip,gisin2007quantum}, distributed quantum sensing\cite{proctor2017networked,proctor2018multiparameter,zhang2021distributed}, and interconnecting future quantum computers\cite{gottesman1999demonstrating,cacciapuoti2019quantum,caleffi2024distributed}. Ultra-sensitive and error-free single photon detectors are essential components in implementing such practical quantum networks.
Today's single photon detection system for quantum communication relies on timing electronics \cite{tagger} to extract quantum information
by recording the photon arrival times and counting the number of photons. Superconducting nanowire single-photon detectors (SNSPDs), featured with 
high detection efficiency ($>90\%$), fast reset times ($<1$ns), and picosecond timing jitter, represent the state-of-art quantum detectors to date. 

\begin{figure}[t!]
\centering
\includegraphics[width=0.8\linewidth]{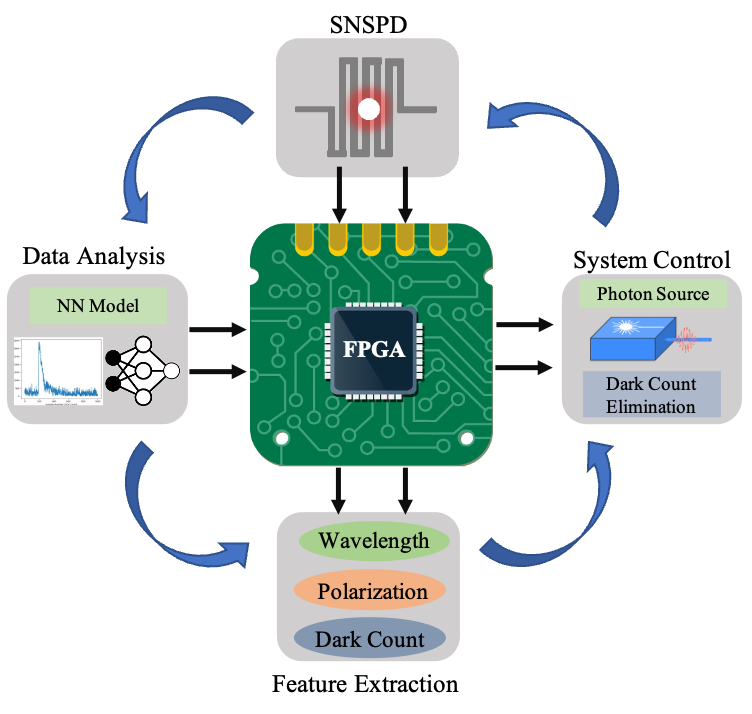}
\caption{An overview of the proposed \name ~system, FPGA is the central processor for data acquisition from SNSPD, data analysis with neural network model, different feature extraction of photons and whole system control.}
\label{fig:overview_design}
\end{figure}

However, most of today's single photon detection system has limited functionality of counting the number of photons. Other physical characteristics of individual photons such as wavelength and polarization cannot be directly detected by the current single photon detection systems. Additionally, almost all the photon detectors operate with non-zero dark counts - false positive clicks in the absence of input photons. Any dark count, either caused by stray photons from the environment or noise in the electronic circuits, would lead to erroneous detection and inaccurate readout of quantum information. This will degrade the fidelity of heralded entanglement generation between remote quantum memories~\cite{lvovsky2009optical,lei2023quantum,sangouard2011quantum,azuma2023quantum}. To this end, the development of a smart photon detection system that simultaneously allows extraction of multi-dimensional information from single photons and a dark-count-free operation is highly important, and will bring significant advances to quantum communications and networking\cite{li2023q, chaudhary2023learning}, as well as quantum information science at large.

\noindent$\bullet$ \textbf{\name:} To overcome the limitations of current photon detection systems, we realize a smart quantum detection system named \name.
For the first time, this \name system leverages a machine learning model for the multi-dimensional feature recognition of individual photons and real-time dark counts discrimination . The implemented \name system, shown in Figure 1, includes a SNSPD operating at $\approx$100 mK in a cryogenic refrigerator\cite{zu2022development}, along with a field programmable gate array (FPGA) board for data recording, neural network machine learning\cite{srinidhi2021deep}, and dynamic system control. With the \name system, multi-dimensional information in individual photons can be retrieved, which includes photon arrival times with pico-second accuracy, the photon number at the detector input, the wavelengths or polarization states of individual photons, and discrimination against detector dark couns. The \name system is fully integrated and compatible with existing quantum communication hardware, ready for plug-play deployment.

\noindent$\bullet$ \textbf{Intuition:}
The working principle of a SNSPD is to convert the energy of an incoming photon to a localized heating (a hot spot) in the superconducting nanowire, where the growing hot spot temporarily causes the local section of the nanowire to be in a normal resistive state, giving rise to a voltage spike at the SNSPD output. In this microscopic picture, there is a correlation between physical properties of the photons, such as frequency or polarization, and the SNSPD voltage response due to the local hot spot formation. For a photon with a specific energy $E = hf$, where $h, f$ represents the plank constant and frequency of the photon \cite{gerry2023introductory}, we expect a uniquely identifiable voltage waveform. This photon-energy to voltage waveform mapping can be exploited to discriminate dark counts\cite{caputo2021dark,vyhnalek2020single}, which are typically caused by either a stray long-wavelength (lower frequency) photons from the environment or spurious electrical noise in the SNSPD circuit itself. By classifying voltage waveforms with respect to the input photon properties, we can therefore establish a data-driven method for extraction of multi-dimensional properties of a photon from the SNSPD output.

\noindent$\bullet$ \textbf{Contributions and results:} The main contributions of \name are: 

1) To our knowledge, \name is the first smart quantum detection system that integrates a fully connected neural network model with an FPGA for multi-dimensional information analysis for photons. 

2) We provide a detailed data acquisition method and develop a customized neural network-based classification model with a fully connected layer that can a) recognize different photon features and b) discriminate dark counts from real photons detected by the SNSPD.

3) We build a fully functional quantum detection system for the performance testing of \name, and demonstrate up to 100\% detection accuracy.

4) We deploy and evaluate the practical performance of the \name to a quantum emitter source based on single erbium ions.

The results show that our \name system achieved up to 100\% accuracy in detecting the wavelength and polarization of single photons, and discriminating dark count errors. By using the \name system, we also show a significant performance improvement in readout fidelity of a quantum emitter.

\section{Background and Challenge Statement}

This section will give a background review of the current single photon detection technology and the potential challenges of the proposed smart quantum detection system. 
\begin{figure}[h!]
    \subfloat[]{
    \includegraphics[width=0.4\linewidth]{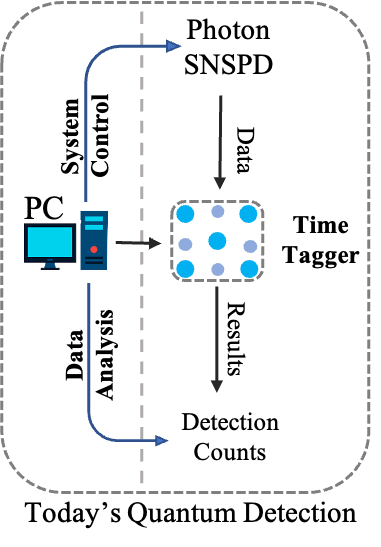}
    \label{subfig:time_tagger}
    }  
    \hspace{0.3 cm}
    \subfloat[]{
    \includegraphics[width=0.44\linewidth]{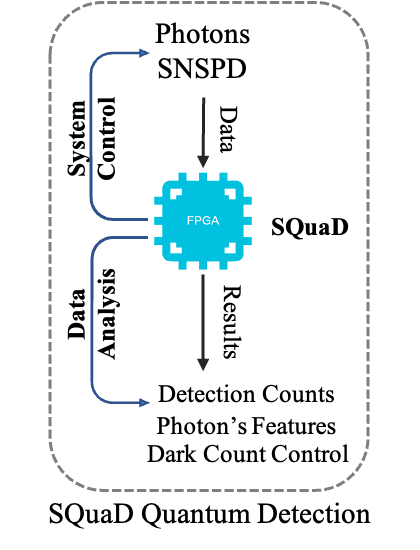}%
    \label{subfig:squad}
    }
    \caption{ Comparison between the current single photon quantum detection system with our proposed smart quantum detection system-SQuaD.}
    \label{fig:comparison}
\end{figure}

\subsection{State-of-the-Art Photon Detection systems} A typical photon detection system today consists of a single photon detector (e.g. SNSPD) connected to a time-digital converter (TDC) device for recording the photon arrival times and counting the number of the photons. The photon counts and timing information is sent from the TDC to a computer for further data processing. The computer is also often used to provide timing synchronization for the TDC as well as control of the SNSPD. Such a system typically delivers a picosecond level timing resolution, which enables practical quantum applications such as quantum key distribution\cite{terhaar2023ultrafast}, entanglement distribution \cite{chapman2022hyperentangled} over quantum networks. Figure\ref{subfig:time_tagger} shows the simplified architecture of current quantum detection systems. The TDC in such as system acts as a passive component that records count rate and timing information of input photons, and is managed by a computer. It is worth noting that the primary function of current photon detection system is  counting photons by setting a threshold on the SNSPD output signal. The system currently does not allow recording and analysis of the full waveforms of the SNSPD output. Passive operation, requiring separate control, and limited ability of data recording makes the current photon detection systems difficult to adapt to complex quantum information processing tasks.  

In contrast, the \name smart quantum detection system centralizes the system control, data recording, and data analysis in one FPGA. The comparison between the  \name and the conventional quantum detection system is illustrated in Figure\ref{fig:comparison}. As shown in Figure\ref{subfig:squad}, \name obviates the necessity for supplementary hardware components, such as an additional computer or TDC, thus facilitating a more integrated and seamless control of the system. Crucially, our methodology with \name harnesses advanced data analytics to extract significantly more information from input photons, as we will describe in details later.  

\subsection{Problem and Challenge} 
In this part, we explain the challenges in achieving our proposal, \name, from a practical quantum communication perspective.
As demonstrated in Figure\ref{fig:comparison}, the proposed \name needs to precisely capture the data flow, accurate and advanced data analysis, and effectively system control platform.  In summary, the challenges stem from the stringent requirements for advanced hardware, a functional software platform, and an intelligent data processing approach.

\noindent{\bf Challenge 1-Hardware:}
Specifically, the hardware system must possess high temporal resolution to adequately capture the rapidly changing data from quantum detections performed by SNSPD. 

\noindent{\bf Challenge 2-Software Platform:}
The software must function as a comprehensive platform capable of coordinating system control across various hardware devices, managing data acquisition, and facilitating data analysis. 

\noindent{\bf Challenge 3-Intelligent Method:} An intelligent data processing methodology is imperative, designed to precisely capture subtle data variations and distinguish the characteristics of each quantum detection event. Considerations must include background noise processing, the development of effective and compatible data analysis models, recognition of distinct features in each detection, and meticulous management of dark count classification and elimination. Developing effective methodology serves as the cornerstone of \name; seamlessly integrating hardware and software is the key to establishing the functional system.

\section{System Design}
This section first presents the operational architecture of the system as outlined in a functional diagram. Next, we describe the electronic hardware specification and software platform, concluding with a discussion of the SNSPD circuit design. 

\subsection{Functional Block of \name} 
\begin{figure}[h!] 
\centering
\includegraphics[width=1.0\linewidth]{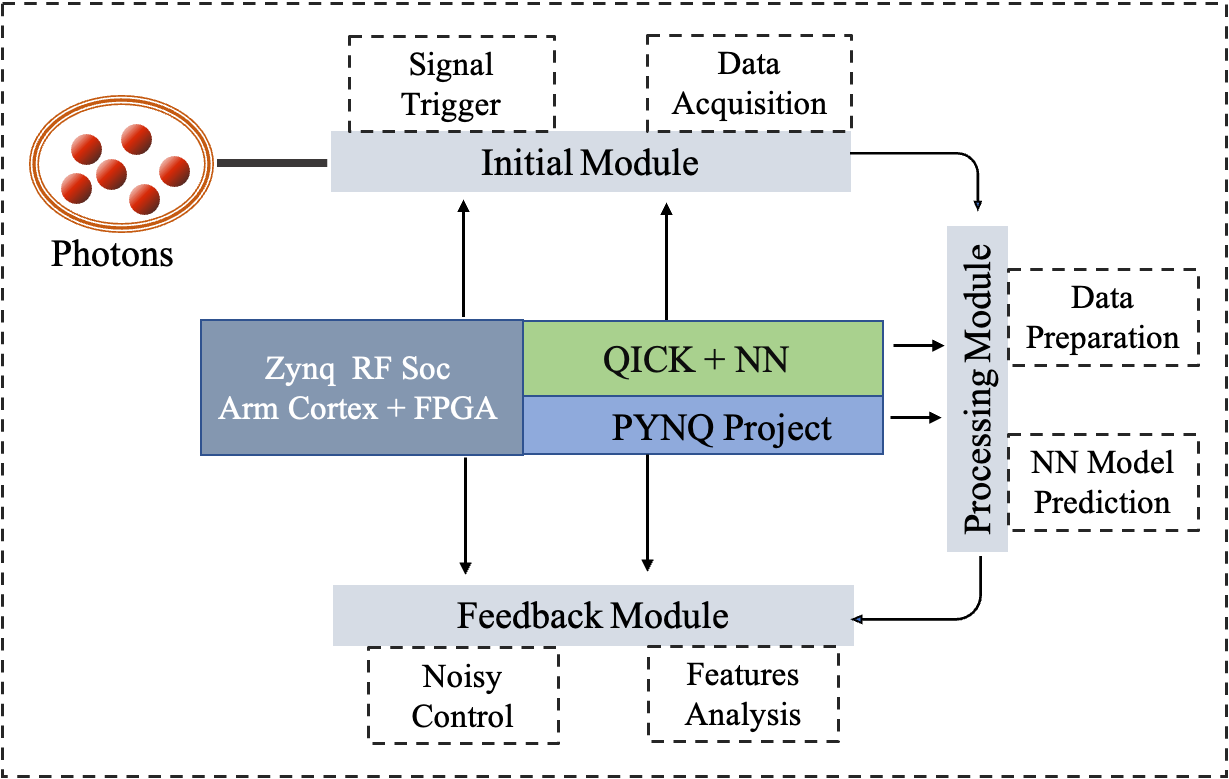}
\caption{Block diagram of \name showing three modules: initial, process, and feedback module, respectively.}
\label{fig:diagram}
\end{figure}

The operational architecture of  \name 
is depicted as block diagrams in Figure~\ref{fig:diagram}, 
which consists of three modules of operation: initial, processing, and feedback module, respectively.

In particular, the initial module provides the signal trigger and data acquisition, where the DAC generates a voltage trigger signal to initiate photon propagation while the ADCs concurrently collect detection data from the SNSPDs.
The collected data of each detection are temporarily stored in the RAM associated with the ARM CPU processor for the usage of next module. The process module employ essistional data preparation as the input of customized NN model. The results of the customized NN model give the predicted feature of each detection. Those features, such as photon wavelength, polarization or dark counts, provide the benchmarks for the feedback module. Dark count elimination or further quantum information related processing will be conducted in this module.

\subsection{Hardware Specification and Software Platform}
This section provides a detailed overview of the hardware specifications and software platform that support the proposed \name system.

\begin{figure}[h!] 
\centering
\includegraphics[width=0.9\linewidth]{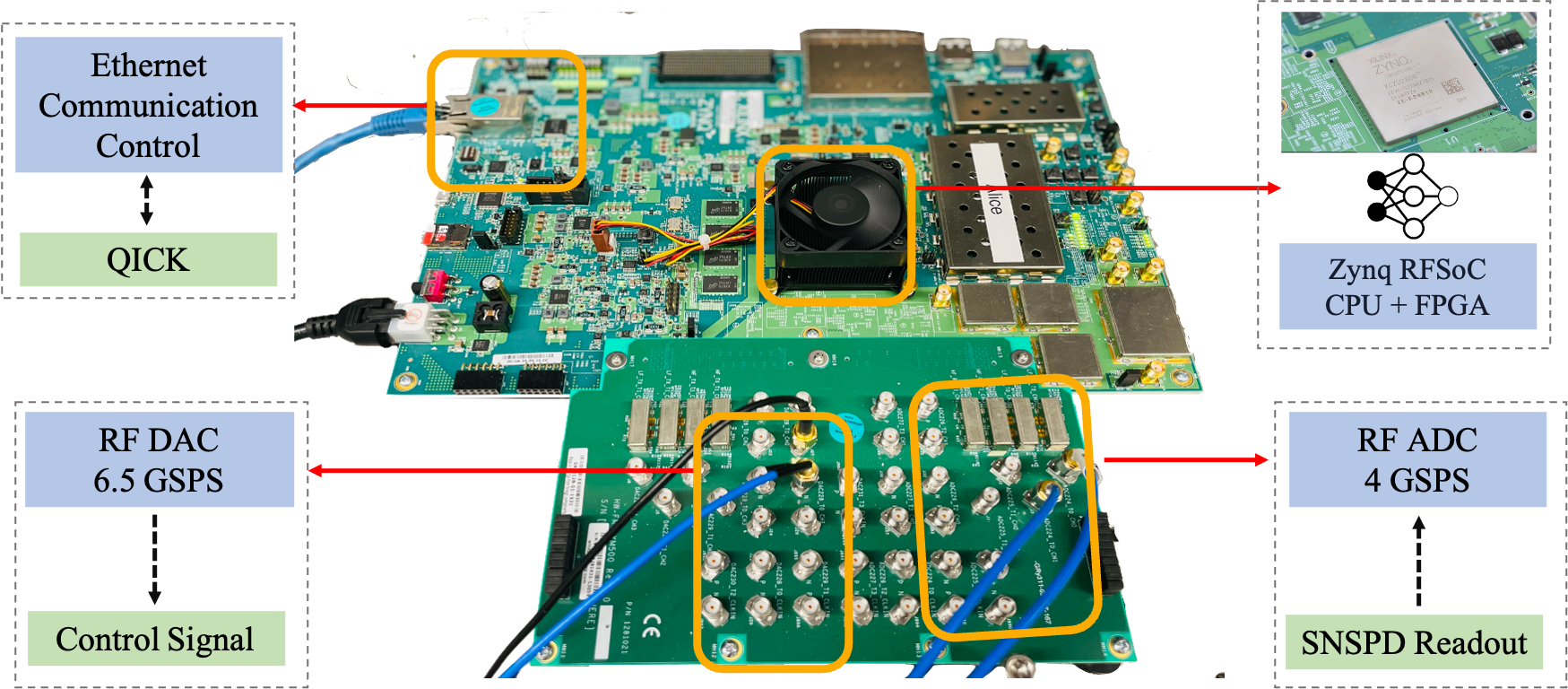}
\caption{ZCU 111 evaluation board for \name system. Hardware specification includes: Zynq RFSoc processing system with CPU + FPGA, RF ADC with 4 GSPS, RF DAC with 6.5 GSPS, and Ethernet ports for data communication and system control}
\label{fig:zcu111}
\end{figure}

{\bf Hardware System Specifications} Figure \ref{fig:zcu111}
depicts the integration of key components on the ZCU 111 RFSoC FPGA board utilized in the \name system. 
For photon detection readout, the \name system employs an RF analog-to-digital converter (ADC), which captures signals from superconducting nanowire single-photon detector (SNSPD) through an electronic cable at a sampling rate of 4 GHz. Concurrently, an RF digital-to-analog converter (DAC) with a maximum sampling rate of 6.5 GHz generates the trigger signal that initiates photon propagation by controlling the operation of the laser pulse. Both the ADC and DAC operations are coordinated by the Zynq RFSoC processor, which integrates an ARM Cortex CPU with a field-programmable gate array (FPGA). The ZCU evaluation board establishes communication channel with the local computer via an Ethernet cable, enabling user interaction for system programming and control.

{\bf Software Platform} As shown in Figure\ref{fig:diagram}, the QICK~\cite{qick} system, integrated with a customized fully connected neural network model operated on the PYNQ framework, serves as the central control management hub in our work. Here, PYNQ is a Python-based open-source software framework, which operates on the ARM Cortex-A53 quad-core processor of Zynq RFSoC architecture integrated on the ZCU 111 evaluation board. 
In particular, the QICK system provides a Python-based programming interface that can customize the operation of FPGA, which includes highly accurate signal generation, data acquisition, and data processing. The proposed \name system integrates the advanced NN models with the QICK system for more sophisticated quantum processing.

\subsection{SNSPD Circuit Design} 
\begin{figure}[t]
    \subfloat[]{
    \includegraphics[width=0.35\linewidth]{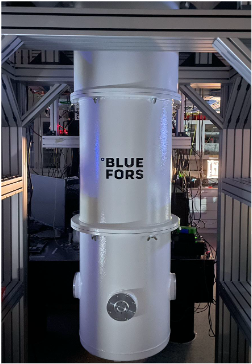}
    \label{subfig:blue_fors}
    }  
    \hspace{0.7cm}
    \subfloat[]{
    \includegraphics[width=0.362\linewidth]{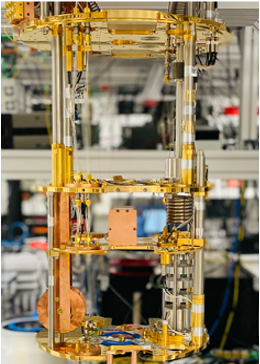}%
    \label{subfig:snsps_dilute}
    }

    \vspace{-0.3cm}
    \subfloat[]{
    \includegraphics[width=0.9\linewidth]{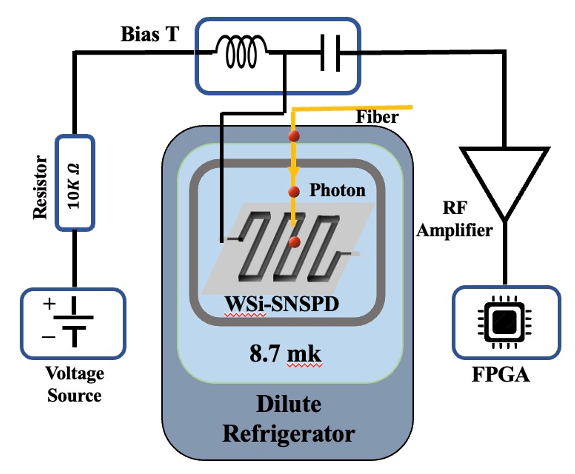}
    \label{subfig:snsps_circuit}
    } 
    \caption{WSi-Based Superconducting Nanowire Single-Photon Detector, the experimental dilute fridge with customized SNSPD depicted in (a) and (b), (c) the electronic circuit for the operation of SNSPD.}
    \label{fig:snspd_all}
\end{figure}

In this experiment, the proposed \name detection system employs an amorphous WSi-based SNSPD nanowire~\cite{oripov2023superconducting} with associated circuit design for the operation of SNSPD.  The cryogenic dilution refrigerator~\cite{bluefors}, depicted in Figure \ref{subfig:blue_fors}, is employed to maintain an ultra-low temperature of 8.7 mK. Within this setup, the WSi-SNSPD is positioned as indicated by the red block in Figure\ref{subfig:snsps_dilute}.  At such low temperatures, the WSi material remains in a superconducting state exhibiting zero resistance, a crucial factor that enables the conversion of the superconducting nanowire into a sensitive single-photon detector. 
To generate a measurable signal, an isolated DC voltage source with a serial connected $10\Omega$ resistor provides a biased current at the mili-ampere level. The biased current flows through a bias tee, creating a closed loop with the SNSPD nanowire. The current variation of the SNSPD due to the photon absorption flow through bias tee, augmented by RF amplifier, and then recorded by FPGA. Concurrently, the integrated SNSPD and FPGA data acquisition system supply raw data essential for training and testing the NN models. The subsequent section will provide a detailed analysis of the SNSPD's working mechanism, complete with theoretical insights and optimal performance. In our work, the voltage source is provided by SRS SIM 928 isolated voltage supply, bias tee is ZFBT-4R2GW, and RF amplifier ZFL-1000LN.

\section{Theory and Performance Analysis of SNSPD}
In this section, we will illustrate the fundamental theoretical analysis of the SNSPD working principle and the associated optimization of SNSPD performance. 
\label{subsec:SNSPD}
\subsection{Theoretical Analysis of SNSPD}
\begin{figure}[t!]
    \subfloat[]{
    \includegraphics[width=0.94\linewidth]{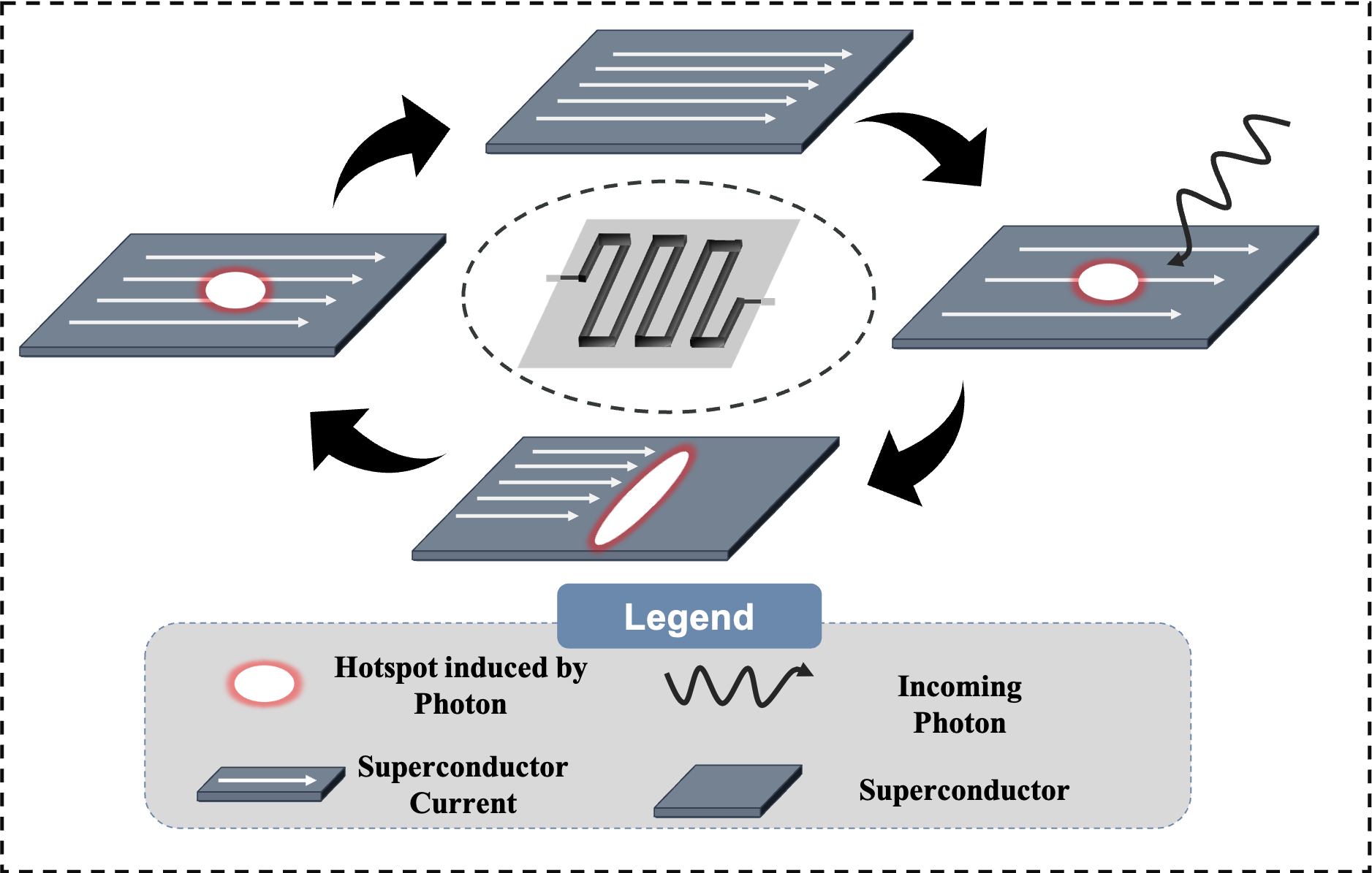}
    \label{subfig:fig:s_con}
    } 
    \vspace{-0.5cm}
    \subfloat[]{
    \includegraphics[width=0.42\linewidth]{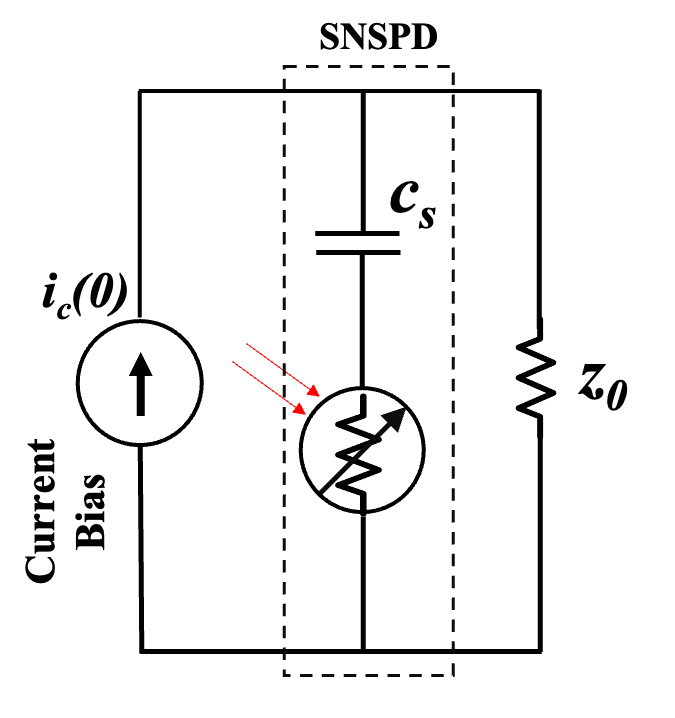}
    \label{subfig:sc_circuit}
    }  
    \subfloat[]{
    \includegraphics[width=0.5\linewidth]{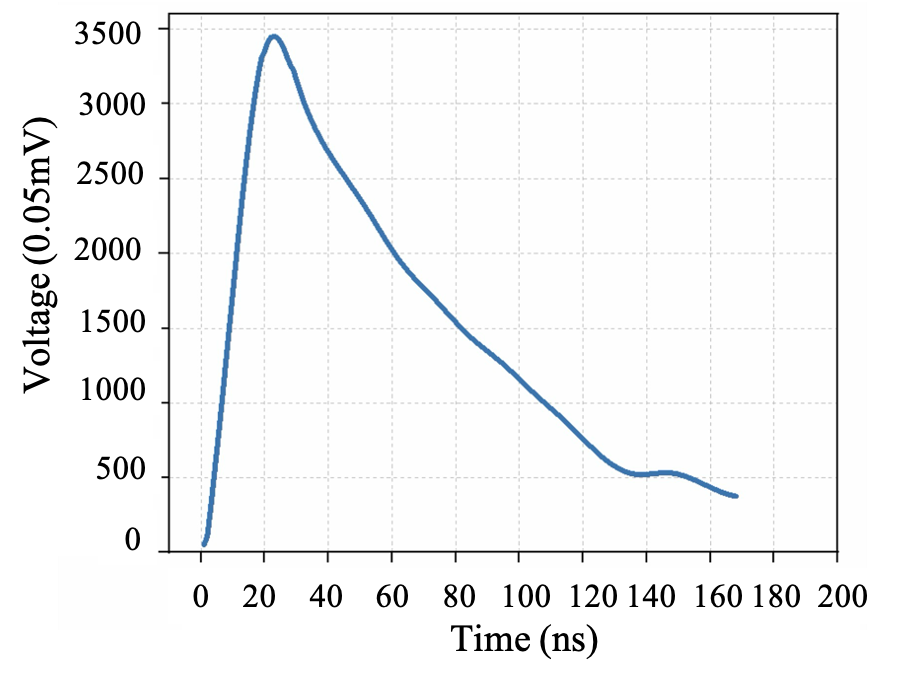}%
    \label{subfig:sc_voltage}
    }
    \caption{(a) SNSPD working principle; (b) Simplified circuit for SNSPD; (c) Voltage change with time from the outcomes of SNSPD.}
    \label{fig:sc_circuit_voltage}
\end{figure}

Fig.~\ref{subfig:fig:s_con} demonstrates the operational mechanism of a superconducting nanowire single-photon detector (SNSPD). 
The zigzag shape sheet represents the WSi superconductor, which is maintained at an 8.7mk dilute refrigerator to preserve the superconductivity.
As explained, the nanowire is connected to a biased current source, forming a closed circuit loop,
where the white arrows represent the current flow circulating through the superconductor nanowire. 
The coming photons interacting with the nanowire will break the superconducting state and go to a normal resistive state, which provides the approach to capture the feature of absorbed photons. 
In particular, the process of transition between superconducting state and the resistive state, as well as the measurement  output in the SNSPD, can be depicted by four phases: {\em absorption}, {\em conversion}, {\em blocking}, and {\em recovery}, which is depicted in Figure\ref{subfig:fig:s_con}. 
During the {\em absorption} phase, the superconductor nanowire traps the photon and absorbs its energy. In the {\em conversion} phase, the absorbed energy is converted into heat, resulting in a localized hotspot with a higher temperature than the surrounding environment on the surface of the superconductor, depicted as the red spot in Fig.~\ref{subfig:fig:s_con}. 
In the {\em blocking} phase, the hotspot expands to cover the entire surface of the superconductor, a resistive barrier is formed, and the current flows are blocked, where the current change of the transition from superconducting state to resistive state is measured in the form of voltage.  
Finally, during the {\em recovery} phase, the heat gradually dissipates into the substrate beneath the superconductor, allowing the temperature to return to its original state, preparing it for the absorption of the next photon.
Mathematically, when considering a photon with time-independent energy, the temperature change in the superconductor can be expressed as~equation \ref{equ:tem_sc_1} and \ref{equ:tem_sc_2} : 

\begin{equation}
     C_{e}\frac{dT_{e}}{dt} =  -\frac{C_{e}}{\tau_{e-p}}(T_{e}-T_{p})+P(t) 
\label{equ:tem_sc_1}
\end{equation}

\begin{equation}
     C_{p}\frac{dT_{p}}{dt} =  -\frac{C_{e}}{\tau_{e-p}}(T_{e}-T_{p})-\frac{C_{p}}{\tau_{es}}(T_{p}-T_{0})
\label{equ:tem_sc_2}
\end{equation}
Equations \ref{equ:tem_sc_1} and \ref{equ:tem_sc_2} illustrate the correlation between the absorbed photon energy and the temperature change of the superconductor. The terms $C_{e}$ and $C_{p}$ provide the electron and phonon-specific heat\cite{natarajan2012superconducting} in this equation, $T_{e}$ and $T_{p}$ are the time-dependent effective temperatures, which can be obtained as a solution of coupled linear heat-balance equations considering the unit volume of film~\cite{semenov1995analysis}, $T_{0}$ is the substrate temperature, $\tau_{e-p}$ is an average electron-phonon interaction time, $\tau_{es}$ is the time of phonon escape from the film into the substrate. $P(t)$ is the time-dependent power of radiation absorbed in the unit volume of the film, expressed as equation \ref{equ:power_radia} to provide the simplest analytical solution of equations \ref{equ:tem_sc_1} and \ref{equ:tem_sc_2}. 
\begin{equation}
    P(t) = \gamma E_{f} m^{3}(2d\tau)^{-1}\xi^{2}exp(-n\xi)
\label{equ:power_radia}
\end{equation}
where m, $\xi = t/\tau$ is the dimension less time, $\tau$ is the full width of the pulse at the half maximum, $\gamma$ is the absorption of the superconductor film, and $d $ is the film thickness. $E_{f}$ is radiation fluence per pulse reaching the superconductor film\cite{semenov1995analysis}, calculated as the ratio between photon energy and coverage area, then amplitude change along the distance is the same as the photon energy change with frequency.   
As mentioned earlier, the temperature change results in the current instability in the superconductor's surface by changing the resistance of the superconductor. 

To have a better understanding of the working principle of the SNSPD, equivalent electric circuit (shown in Fig. \ref{subfig:sc_circuit}) is shown, which responds to resistance change by measuring the voltage change across the load resistor is shown in Fig. \ref{subfig:sc_circuit}. 

Given the biased current in the SNSPD device, the voltage change of the output of SNSPD for each absorbed qubit can be expressed as: 
\begin{equation}
    v = \left(i_{c}(0)-i_{c}(0)\left(1-\left(\frac{T_{SNSPD}}{T_{c}}\right)^{2}\right)^{2}\right)z_{0},
\label{equ:sc_voltage}
\end{equation}
where $i_{c}(0)$ is the biased current connected to SNSPD, $T_{c}$ is the critical temperature of the superconductor film, $z_{0}$ is the load resistor, $T_{SNSPD} =  T_{e}$ is the temperature change of the superconductor film. 

Equations \ref{equ:tem_sc_1} to \ref{equ:sc_voltage} model the operational mechanism of SNSPD  with the photon detection, temperature and superconductivity variation, and voltage output. Most importantly, this model reveals the correlation between photon energy and the voltage output. Fig. \ref{subfig:sc_voltage} depicts the voltage change over time of one complete photon detection. It can be observed that the voltage quickly rises and then gradually returns to normal over time, reflecting the corresponding temperature changes associated with the interaction between the photon and the superconductor.

\subsection{Optimization of SNSPD Performance}
\begin{figure}[h!]
    \centering
    \includegraphics[width=0.8\linewidth]{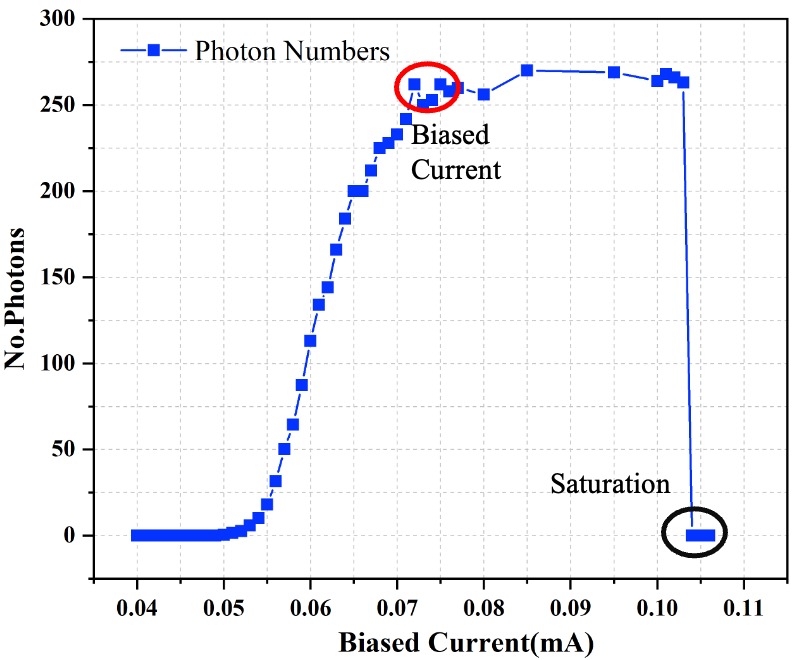}
    \caption{Number of detected photons against the biased current of SNSPD, biased current of 0.07 mA is selected for the \name system evaluation in this work}
    \label{fig:SNSPD_current}
\end{figure}

This section demonstrates the optimization of SNSPD performance to find out the optimal biased current for the readout, which will analyze the optimal state of the SNSPD that satisfies the following requirement: 1) effective photon or dark-count reaction where the fast recovering time is required; 2) the optimal magnitude of SNSPD signal readout which ensures the dataset collection for the following NN model training and testing. As explained in Section 4.2, the working mechanism of SNSPD is equivalent to the LR circuit. Therefore, the magnitude of input current (biased current) is the parameter that significantly affects the performance of SNSPD. In the experiment, the biased current source is provided by SRS600 manufactured by Stanford Research System, where mA of biased current is circulated through the SNSPD nanowire. The number of photons is recorded by the Time Tagger device, manufactured by Swabian. The laser source has a wavelength of 1535nm and directly shines on SNSPD through fiber. To avoid the saturation of SNSPD caused by too many photons approaching, the power attenuator is applied to reduce the number of photons of laser per second. The metric of the performance of SNSPD is characterized by the received number of photons per second under the same amount of power of laser against the variation of the biased current. Figure\ref{fig:SNSPD_current} plots the number of received photons and the biased current change. As shown in the plot, it can be observed that the detected number of photons will increase exponentially, remain constant, and drop directly to zero. The red circle is choosen as the optimal biased current used in this work, and black circle represents the biased current that satureation occurs. This observation manifests that biased current exists at the maximum number of the detected phone, which is labeled as a circle in the plot. In the following experiment, the biased current value that yields the maximum number of photons is identified as the optimal current and is used to collect data for training the NN model.

\subsection{Statistical Analysis and Intuition}
\begin{figure}[h]
    \centering
    \includegraphics[width=0.8\linewidth]{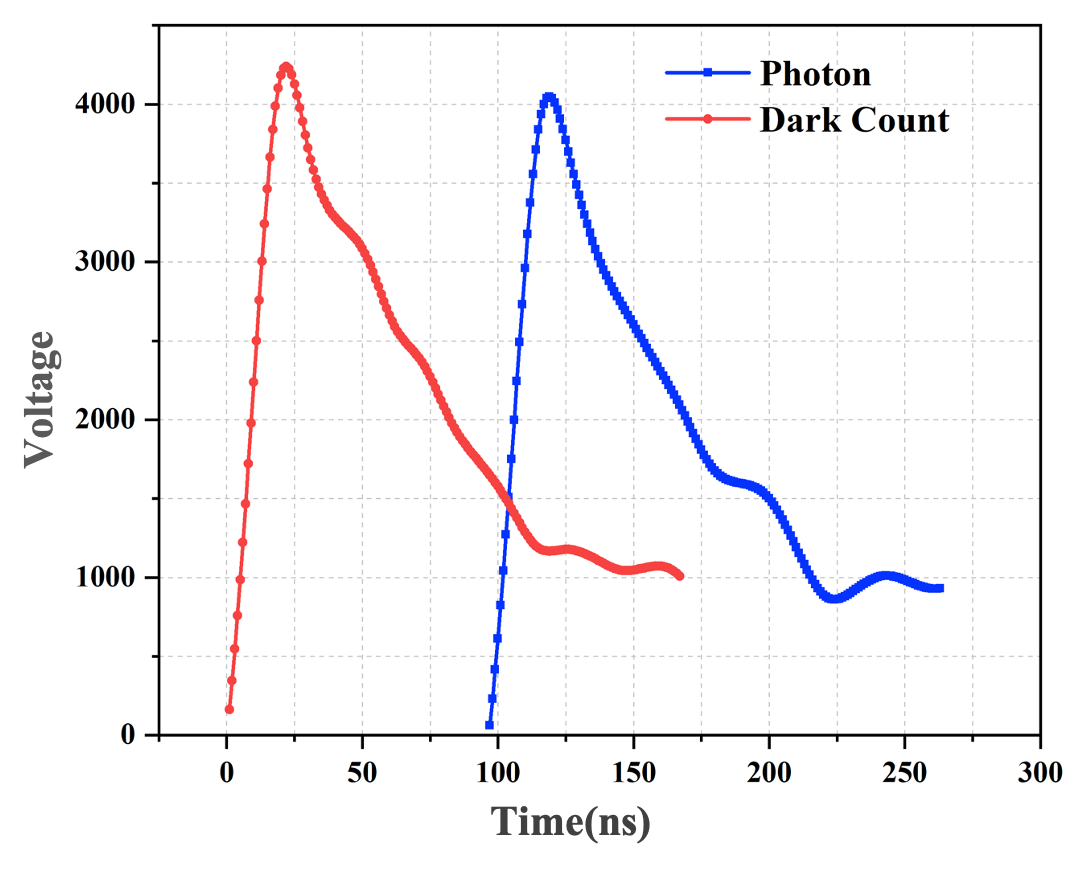}
    \caption{Waveform Comparison for detected photon(at 1535nm) and dark count }
    \label{fig:waveform}
\end{figure}

With the optimal biased current of SNSPD, the simple comparison between photon and dark count is depicted in Figure\ref{fig:waveform}. This figure 
From the obtained figure, the following statistical observations are obtained: 1) both photon and dark count share a similar waveform shape, rising sharply from the basis and going back to the basis slowly; 2) the amplitude of photon is smaller than dark count, which indication the hypothesis that the physical features of the dark counts are the unexpected photon from the environment with higher frequency, such as the ultra red light from the room temperature; 3) the complete bandwidth of photon or dark count are around 100 nanoseconds; 4) the fluctuations occurred for both photon and dark count, which could alter the amplitude and bandwidth of the waveform signal. Those observations validate the optimal performance of SNSPD. More importantly, those observations experimental demonstrate the evidence of dark count elimination from normal photons and concrete the intuition of the proposed \name. Those differences observed from this figure is utilized as the input of NN model training and testing. 


\section{Proposed Neural Network-based Models in \name}
\label{sec:proposed_neural_network}
The \name leverages fully connected neural network-based machine learning models (FCNN) for solving real photon and dark count classification problems. First, we describe the dataset collection and features chosen for training the following FCNN models. Then, we explain the methodology for designing the neural network architectures.  

\subsection{Dataset Collection and Notations}
\label{subsec:dataset_notations}
We obtain the dataset as the inputs of the following proposed NN training model from SNSPD and FPGA described in Sec.~\ref{fig:overview_design}. 
The collected data are the raw data from ADC (analog-digital converter). Here, we consider three distinct categories, each corresponding to the cause of influence on the detected photon or dark count of SNSPD. These categories are defined as follows: (1) {\tt Entry}: the features of photon (wavelength, polarization) and dark count  ; (2) {\tt Setting}: parameters of detection devices; and (3) {\tt Deliverable}: the readout of the SNSPD, mentioned in Section 3. 
We collect and combine datasets from various scenarios to support the proposed neural network (NN) model for two main objectives: (i) classification between real photon and dark count and (ii) recognition of different photon features.

\subsection{{\color{black} Neural Network-based Classification Model}}
\label{subsec:classifier}

\noindent{\bf Data matrices for classifier:} We define the data matrices for {\tt Entry}, {\tt Setting} and {\tt Deliverable} as: $X^{\classifier}_{E} \in \mathbb{R}^{N^{\classifier}_t \times d^{\classifier}_{E}}$,  $X^{\classifier}_{S} \in \mathbb{R}^{N^{\classifier}_{t} \times d^{\classifier}_{S}}, \text{ and } X^{\classifier}_{D} \in \mathbb{R}^{N^{\classifier}_{t} \times d^{\classifier}_{D}}$, respectively, where $N^{\classifier}_{t}$ is the number of training samples. Furthermore, $d^{\classifier}_{E}$, $d^{\classifier}_{S}$, and $d^{\classifier}_{D}$ give the dimensionality of {\tt Entry}, {\tt Setting} and {\tt Deliverable} categories. The $|d^{\classifier}_{E}|$, $|d^{\classifier}_{S}|$, and $|d^{\classifier}_{D}|$ are optimized based on the training results of the proposed NN model. The set of the output labels are: $L_{\classifier} = \{\match , \mismatch\}$. 
Here, $\match$ means the accurate prediction of the input qubit state, such as $0 \mapsto 0$ or \textcolor{black}{$1\mapsto 1$},
and $\mismatch$ means inaccurate prediction, such as $0\mapsto 1$ or $1\mapsto 0$. 
Here, we statistically define $0$ and $1$ as dark count and photon, different frequency, or different polarization, respectively.  
We consider the label matrix $Y_{\classifier}\in \{0,1\}^{N^{\classifier}_t\times |{L}_{\classifier}|}$ that represent the one-hot encoding for the dark count and photon prediction, or different features of the photon.

\noindent{\bf Neural network-based model as a classifier:} 
The fully connected neural network used in this model is a combination of different layer-based linear feature extractors followed by a {\em Sigmoid} non-linear activation. We represent the ultimate layer transformations with  $ f_{\theta_{\classifier}}$, parameterized by weight vector $\theta_{\classifier}$. The $ f_{\theta_{\classifier}}$ maps the input to the one-hot encoded output. Formally, we define the ultimate layer of the network as: %
\begin{equation}%
\scores^{\classifier} = \sigma(f_{\theta_{\classifier}}(X^{\classifier}_{E}, X^{\classifier}_{S}, X^{\classifier}_{D})),~~~~~f_{\theta_{\classifier}}:\mathbb{R}^{d^{\classifier}_{E} + d^{\classifier}_{S} + d^{\classifier}_{D}} \mapsto \mathbb{R}^{|Y_{\classifier}|} 
 \label{eq:classifier}%
 \end{equation}
where $\sigma: \mathbb{R}^{|Y_{\classifier}|} \mapsto  (0, 1)^{|Y_{\classifier}|}$ signifies the {\em Softmax} operation, and $ \scores^{\classifier}$ is the prediction score of the network. Overall the prediction of photon or dark count is solved using the neural network by: \( \mathcal{C}  (.)  \) = $\arg\max \sigma(f_{\theta_{\classifier}}(.))$.

\noindent
{\bf Model training: }The learning model $f_{\theta_{\classifier}}(.)$ is a function parameterized by $\theta_{\classifier}$, i.e., a neural network with weights $\theta_{\classifier}$. The empirical loss of the model parameters ${\theta_{\classifier}}$ on the $j^{th}$ sample of the dataset is defined as $ \mathcal{L}({\theta_{\classifier}}, j) :=  [\ell(f_{\theta_{\classifier}}(X^{\classifier}_{E_j}, X^{\classifier}_{S_j}, X^{\classifier}_{D_j}),Y_{\classifier_j})]$, where $\ell$ is a loss function measuring the discrepancy between predicted and true labels, cross entropy as an instance.
The standard DL training approach finds a model that minimizes the loss across all of the training samples by solving:
    $\underset{\theta_{\classifier}} {\min} \frac{1}{N^{\classifier}_{t}} \sum_{j=1}^{N^{\classifier}_{t}} \mathcal{L}({\theta_{\classifier}, j})$. Over the model training, we achieve the optimum values of $\theta_{\classifier}$, which is used for predicting different features of photon/dark count during inference.

\noindent
{\bf Model inference: } After the model training, the output of whether the correct unitary state is predicted or not is inferred from: 
\begin{equation}
    y_{\classifier_l} =  \sigma(f_{\theta_{\classifier}}(X^{\classifier}_{E_l}, X^{\classifier}_{S_l}, X^{\classifier}_{D_l})),
\label{eq:classifier_prediction}
\end{equation}
where $\sigma$ denotes the {\em Softmax} operation,  $X^{\classifier}_{E_l}, X^{\classifier}_{S_l}$, and $X^{\classifier}_{D_l}$ are test samples from {\tt Entry}, {\tt Settings}, and {\tt Deliverable} categories for $l^\text{th}$ samples, respectively.

\begin{figure}[h]
    \centering
    \includegraphics[width=\linewidth]{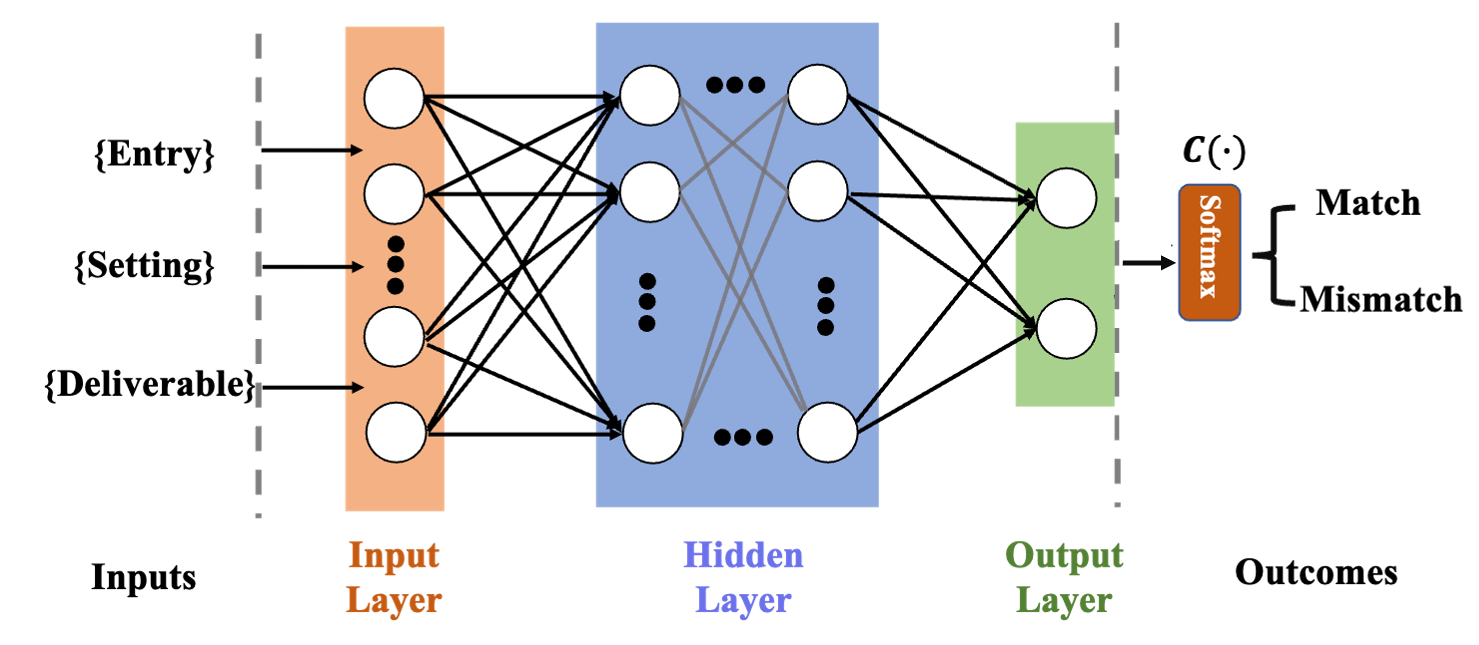}
    \caption{The proposed fully connected neural network architecture with different number of layers and neurons}
    \label{fig:fc_nn}
\end{figure}

The generic idea of the proposed classifier for the {\tt MATCH} or {\tt MISMATCH} detection is presented in Fig.~\ref{fig:fc_nn}.

\section{\name ~Experiments}
\label{sec:experimental_evaluation}
This section demonstrates the experimental results of the proposed \name system. First, we depict the dataset collection for machine learning model training. Then, the methodology and the performance of the neural network model is demonstrated. Three different applications of the proposed \name system are implemented at the end.



\noindent
\textbf{Specification and Evaluation Metrics} Those results are performed on the Python version 3.9.13, using an Intel(R) Core(TM) i7-9750 CPU at 2.60GHz with RAM 32.0 GB machine for machine learning model training, ZCU 111 RFSoc evaluation board for data acquisition through ADC, and machine learning model testing of \name.
We use the {\em testing accuracy} to evaluate the performance of the regression model for the readout of the SNSPD calibration and classification model for photon and dark count.  

\subsection{Dataset Collection for Machine Learning Model Training}
This section illustrates the experimental setup for original data collections and the neural network model (classifier) training/testing performance. 

\noindent{\bf Dataset Collection Setup:} The dataset collection of \name ~in this work consists of photons with different wavelengths, polarization, and dark count for the neural network model training. The wavelength ranges from 1520nm to 1550 nm with up to 1-nanometer variation. The polarization recognition associated with the proposed \name in this work emphasizes vertical and horizontal. Additionally, for the purpose of being compatible with the dark count elimination task on the erbium photon emitter prototype, photon data at a wavelength of 1535nm with the laser locking method are collected, and the dark counts are collected without a laser source in a dark environment. The details of laser locking approach based Pound-Drever-Hall (PDH)methond can be found in ref\cite{black2001introduction}. The photon source, generated by TOPTICA laser light, offers varying wavelengths and polarization. 

\noindent{\bf Photon with Different Wavelength and Polarization:}
For the data set of different wavelengths or polarization,  the generated laser light with photons propagates through the fiber and hits the SNSPD directly. 
The distance between the laser source and SNSPD is 5 meters, and an optical attenuator is applied to reduce the power of the laser light and the number of photons per second to avoid the saturation that happens in the SNSPD. 
In the experiment, we set the photon rate as 10kHz, the collection time is 0.5 seconds, and the total number of photons is 5000 for each wavelength or polarization separately. For the scenario of different wavelengths,  wavelength selection is centered at 1535 nm, 
with variations of  $\pm 1, 2, 5, 10, 15$ nm from 1520nm to 1550 nm. 
Subsequently, the classifier used in the \name is utilized to recognize the difference between the central wavelength and each varied wavelength in this work. 
Regarding the different polarization of photons, a three-pad polarizer is used to tune the polarization of photons installed between the laser source and SNSPD. In particular,  vertical polarization and horizontal polarization are considered in this work, and the wavelength is fixed at 1535nm for both polarizations, respectively.

\noindent{\bf Photon and Dark Count:}
As mentioned above, the data collection for the scenarios of distinguishing photon and dark count utilized laser locking to be compatible with the prototype test of erbium photon emitter dark count elimination. Wavelength with 1535nm of photon is considered here.
In terms of the dark count data collection, the experimental setup is the same as the case of collecting photons, except that the laser is off. During the dark count data collection, the laser source is off, and all of the room lights are off. In our experiments, the dark count rate is 2-3 Hz. Hence, to be identical with the number of photons, the collection time for the dark count is 2400 seconds for each time data set collection.

\subsection{Methodology of Data Analysis and Performance of Machine Learning Model}

This section demonstrates the data analysis methodology for FCNN model training and the associated performance.  It is important to highlight that the FCNN model developed for the \name system focuses on classifying two specific types with the binary models, such as the wavelength between 1535nm and 1540 nm,  vertical polarization and horizontal polarization, or photon and dark count.  While the methodology for classifying additional types remains consistent, the associated performance metrics for these classifications will not be covered in this work. In particular, the methodology of leveraging the machine learning approach to recognize the photon's wavelength and polarization or distinguish the dark count from the photon is the same. Hence, to avoid the repeated narratives of the same knowledge, distinguishing dark count from photons is demonstrated as an example to show the methodology and implementation of the FCNN model approach to the \name . Then, the evaluation performance related to photon wavelength, polarization, and dark count elimination is demonstrated in the following sections.

\begin{figure}[h]
    \centering
    \includegraphics[width=1.0\linewidth]{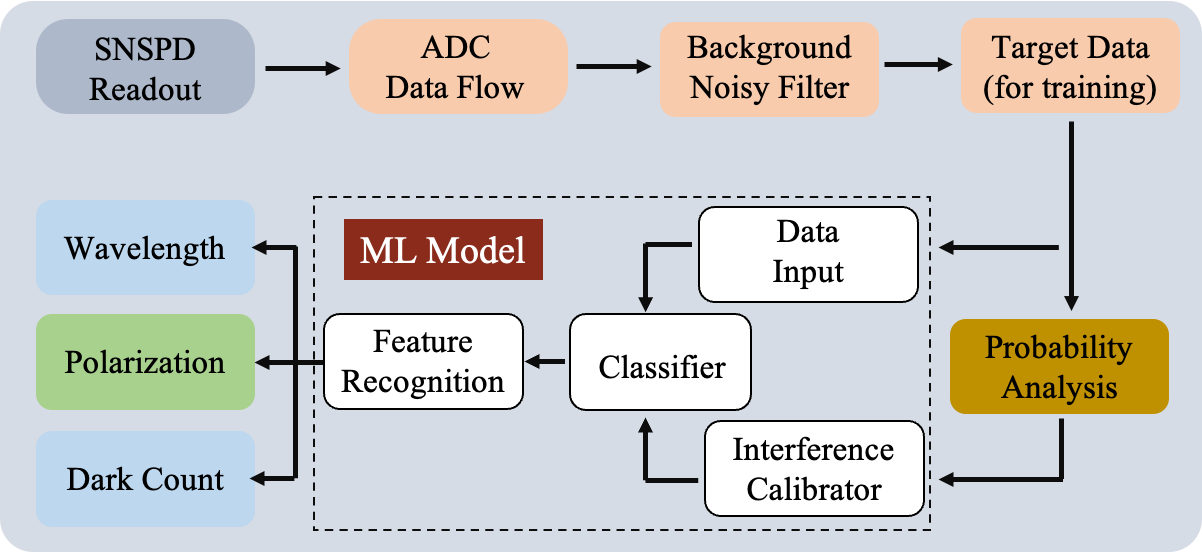}
    \caption{Workflow for training the classifier with the raw data obtained by FPGA }
    \label{fig:work_flow}
\end{figure}

\noindent{\bf Data Processing Workflow:}Figure\ref{fig:work_flow} illustrates the workflow of source data processing with the purpose of machine learning model training. The source data processing consists of background noisy filtering, interference calibration and feature recognition with a classifier model 

\begin{figure}[b]
    \centering
    \subfloat[]{
    \includegraphics[width=0.5\linewidth]{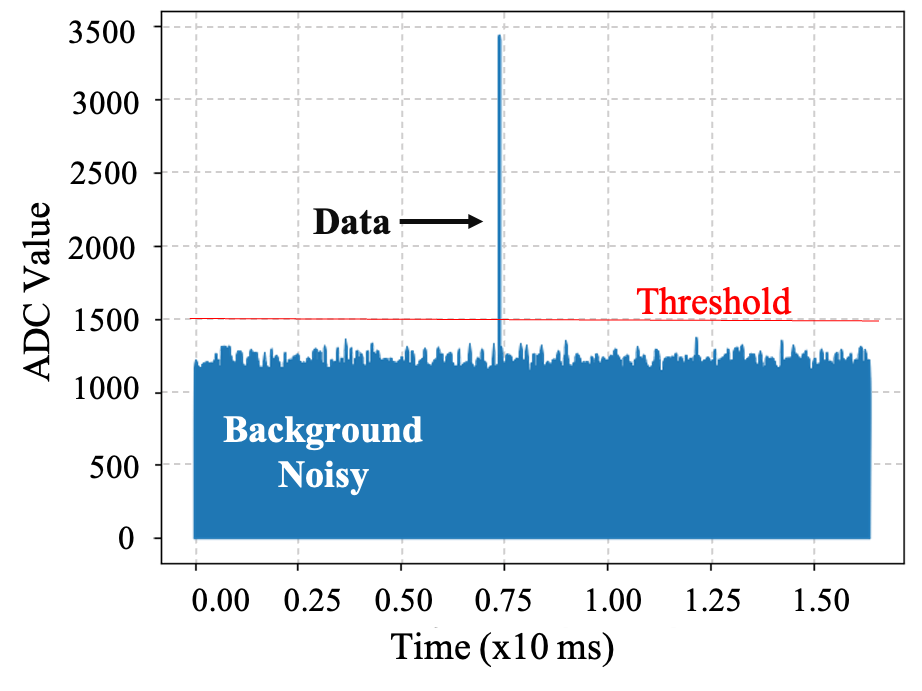}
    \label{subfig:raw_unfiltered}
    }     
    \subfloat[]{
    \includegraphics[width=0.5\linewidth]{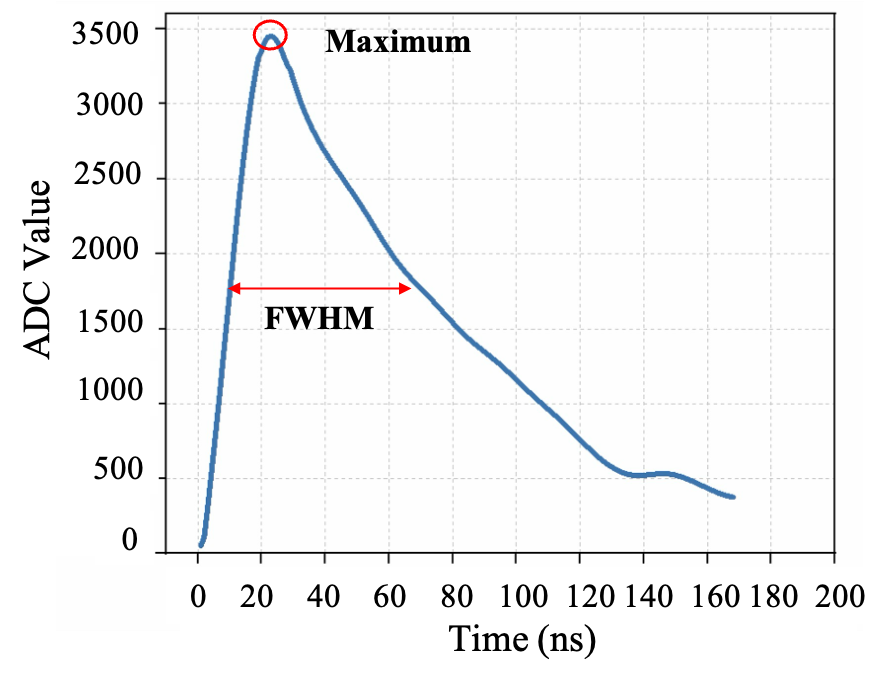}%
    \label{subfig:raw_filtered}
    }
   \caption{Waveform data variation between photon and dark count}
    \label{fig:raw_data_filtering}
\end{figure}

\noindent{\bf Background Noisy Filter:}As illustrated in Figure\ref{fig:work_flow}, the SNSPD readout as the source data flow to ADC. The source data consists of both the background noisy and the photon detection signal, and the source data is plotted in Figure\ref{subfig:raw_unfiltered}. In this figure, the x-axis represents the time duration, and the y-axis stands for the voltage value recorded by the ADC. In this work, the voltage value after ADC is utilized for the FCNN model training. Also, it can observed from Figure\ref{subfig:raw_unfiltered}(a) that a clear boundary exists between background noise and photon detection, and a threshold-based approach is used as the background filter. Figure\ref{subfig:raw_filtered}(b) plots the data filtered from the background noise. The features of the filtered data, maximum value, FWHM, the rising and falling time, are applied as the data input of the following classifier model training.

\begin{figure}[h]
    \centering
    \includegraphics[width=0.9\linewidth]{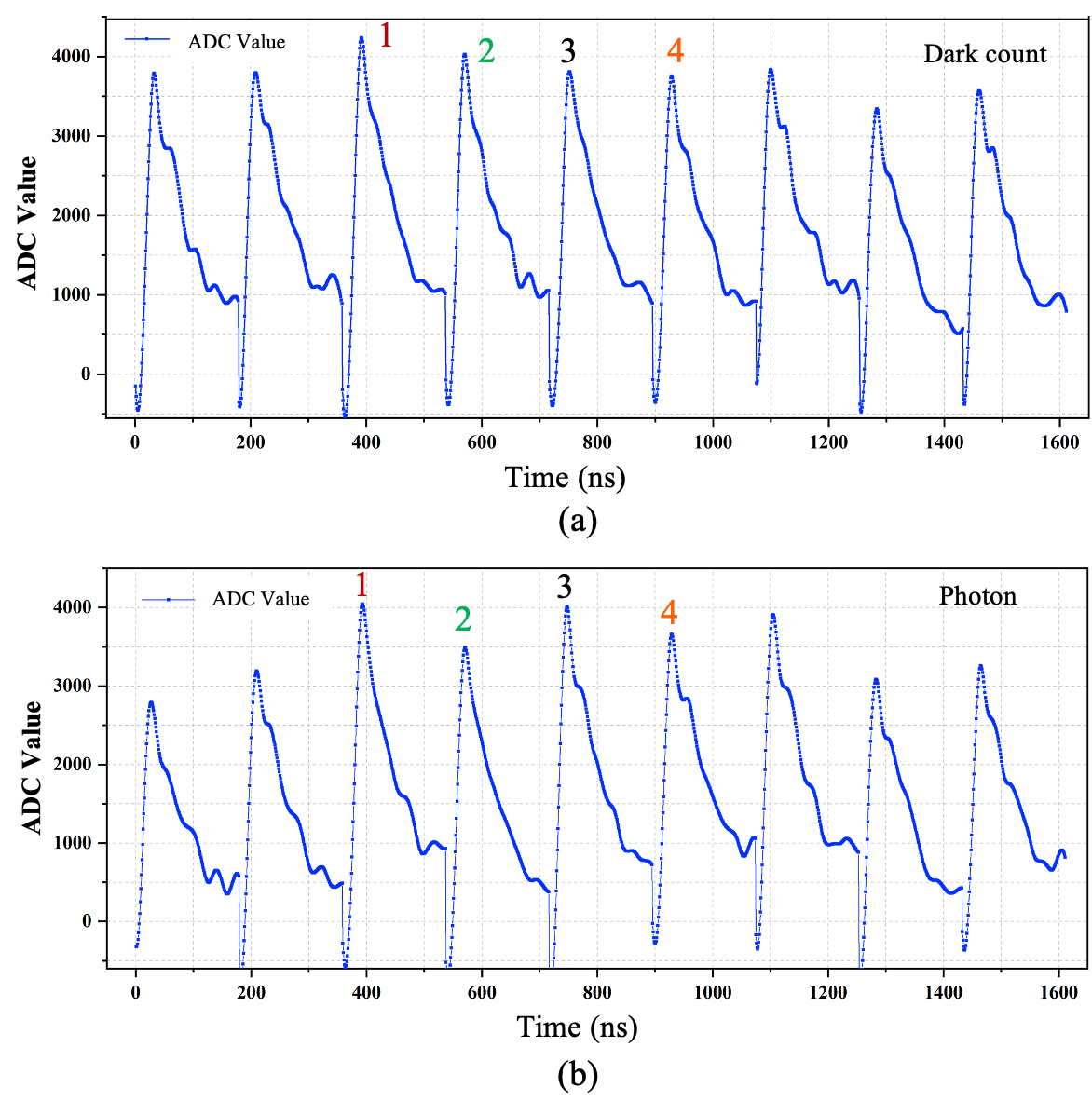}
    \caption{Experimental Setup for photon/dark count collection}
    \label{fig:data_cal}
\end{figure}

The data filtering process removes the noise from the background. In our experiment, the FCNN model directly leverages these values in the training to cla, However, the filter data demonstrates different variations even for the photon with the same physical property. 
Figure \ref{fig:data_cal} plots the waveform variation of the filtered data for photon and dark count, depicted as the labels 1, 2, 3, and 4, respectively. It is observed that the waveform shows different maximum values and the FWHM for only photon or dark count, which shows disagreements with the theory where photon or dark count is supposed to have the same waveform change. In this paper, we have the hypothesis that these differences are deduced from the electronics and minor temperature changes of the SNSPD working environment. Therefore, the interference calibration is conducted on the filtered data, and the associated calibrator is used as the other input parameters for the classifier model training.    

\begin{figure}[t]
    \centering
    \subfloat[]{
    \includegraphics[width=0.5\linewidth]{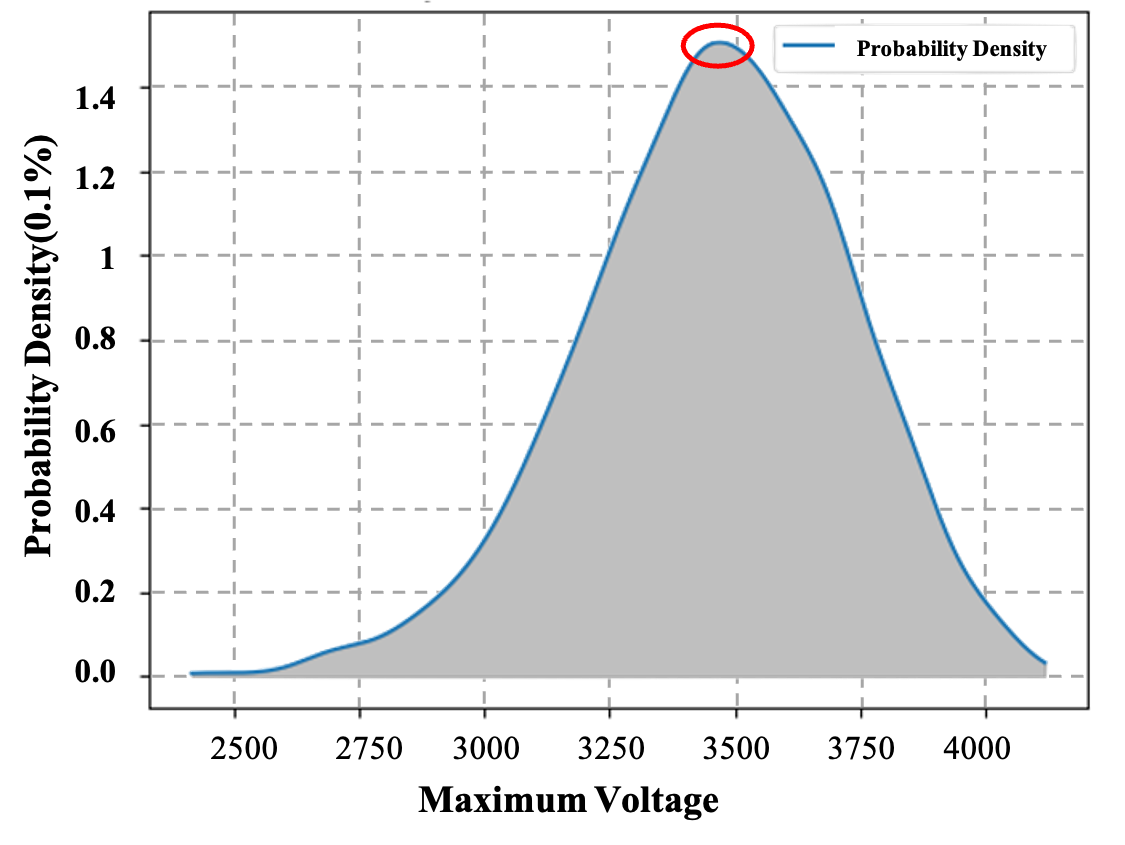}
    \label{subfig_unlock}
    }  
    \subfloat[]{
    \includegraphics[width=0.5\linewidth]{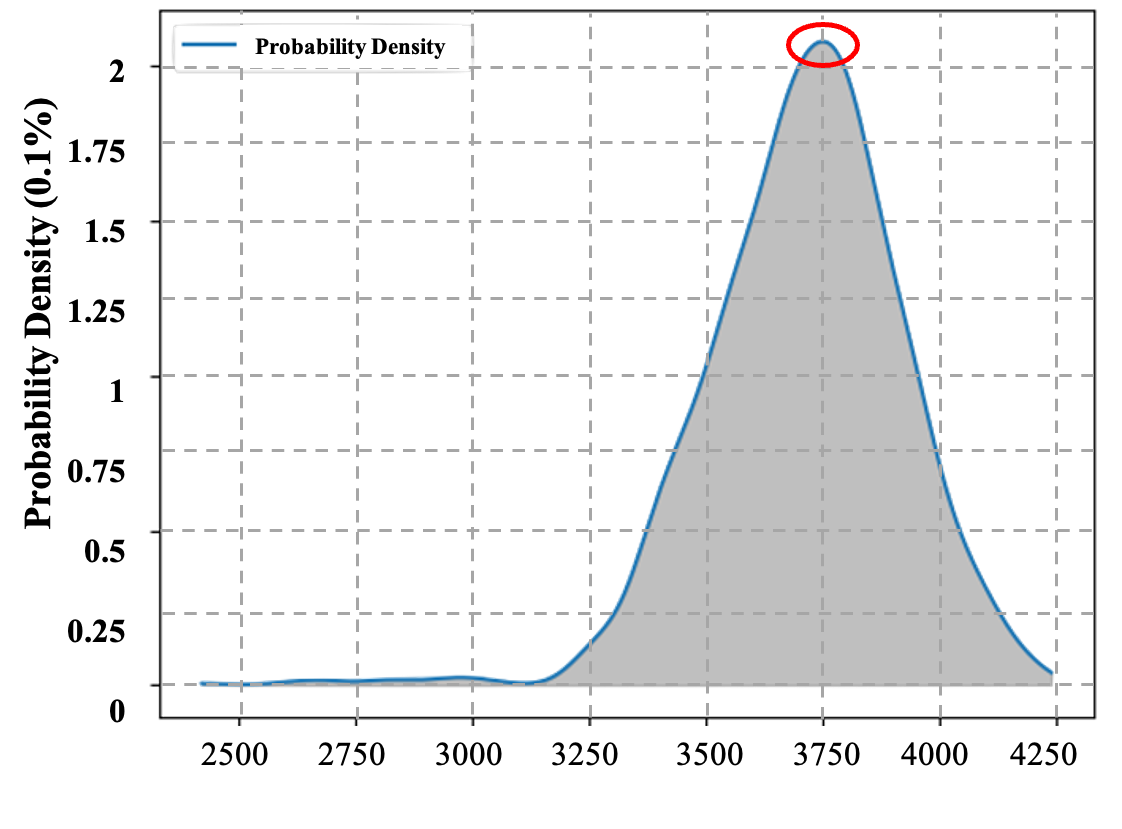}%
    \label{subfig_lock}
    }
   \caption{Distribution histogram of photon and dark count based on the detected possibility}
    \label{fig:data_density}
\end{figure}

To calibrate the data values for photons and dark counts, establishing a reference basis is crucial. This reference is determined by plotting distribution histograms of the maximum values in the collected data for both photons and dark counts, based on their respective probabilities. The histogram plots are shown in Figure\ref{fig:data_density}; it is observed that both the maximum value for photon and dark count demonstrate approximately a normal distribution. Then, the reference basis for the calibration of the photon and dark count is the value at the point where it has maximum possibility, which is lined as red in Figure\ref{fig:data_density}, 3400 for photon and 3800 for dark count. 

Given the probability distritbuion, we define a calibration factor that represents the scale of adjustment, defined as a linear function between probability and the associated maximum value of each detection and expressed as $factor = f(p_{max}, v_{max}$.
With this approach, the input data each detection for the FCNN model training consists of the original voltage value data as shown in Figure\ref{subfig:raw_filtered}   and the calibration factor.  

\begin{figure}[h]
    \centering
    \includegraphics[width=0.8\linewidth]{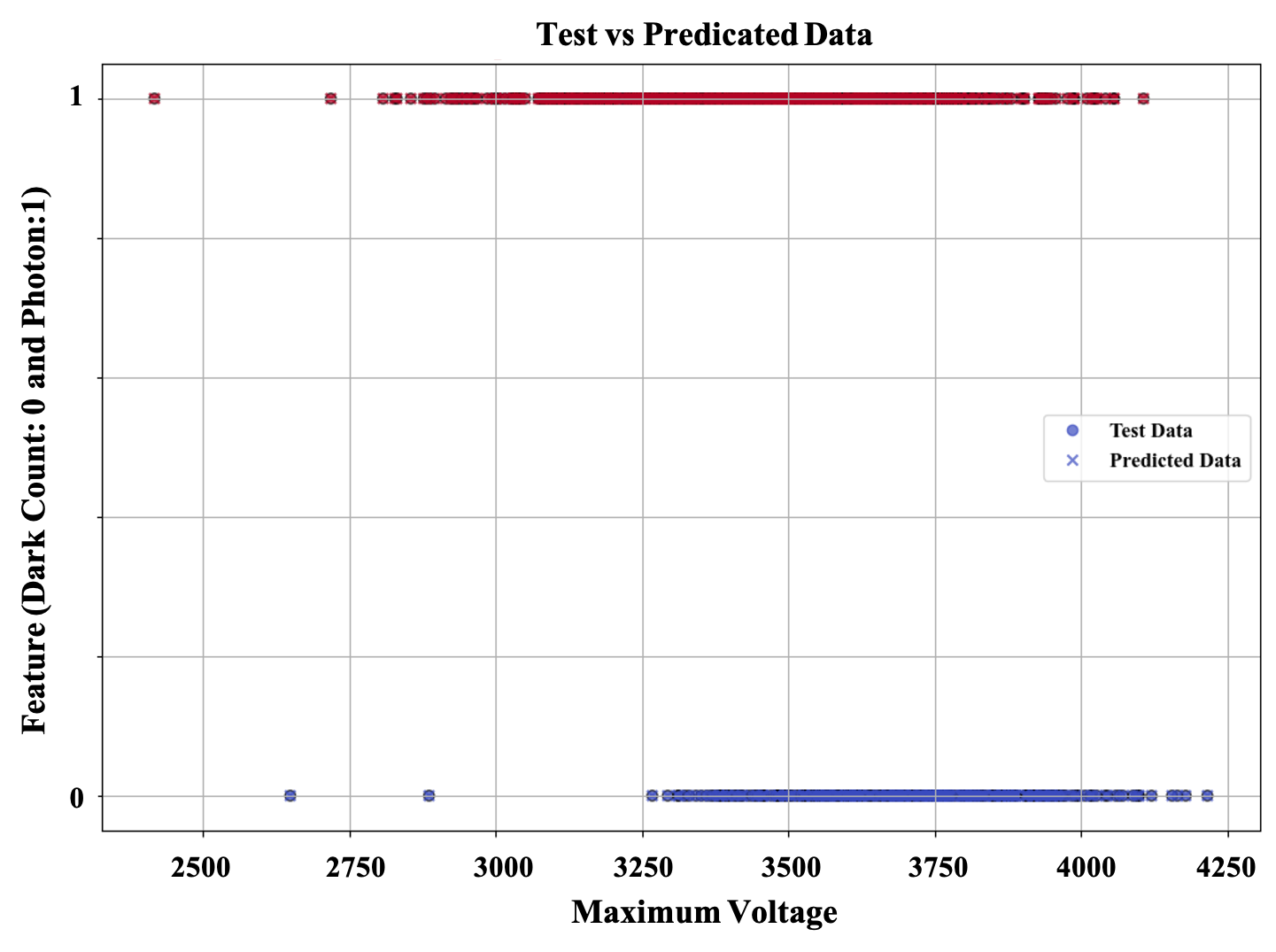}
    \caption{Comparing results of test data and predicted data for the classification between dark counts and normal photons}
    \label{fig:data_class}
\end{figure}

\begin{table}[h]
\centering
\small
\scalebox{0.9}{\begin{tabular}{|c|c|c|c| c |}
\hline
Category & \multicolumn{1}{c|}{\tt Entry} & \multicolumn{1}{c|}{\tt Setting} & \multicolumn{2}{c|}{\tt Delievrable}\\
\hline\hline
 Dark Count & \thead{Dark Counts\\ or Photon} & \thead{Biased \\ Current}  & \thead{Voltage \\ Time Variation} &  \thead{Calibration \\ Factor}  \\
\hline
 Wavelength & \thead{Different\\ Wavelength} &\thead{Biased \\ Current} & \thead{Voltage \\ Time Variation}  & \thead{Calibration \\ Factor}  \\
\hline
 Polarization &\thead{Vertical\\ Horizontal} & \thead{Biased \\ Current} & \thead{Voltage \\ Time Variation} &  \thead{Calibration \\ Factor} \\
\hline
\end{tabular}}
\caption{Data Setting for the FCNN model training and testing.}
\label{tab:input_cases}
\end{table}
At the end of the neural network model training, input data are the integration of the voltage variation with time, and the calibration factor of each detection. As a summary, Table \ref{tab:input_cases} list all of data input used for the FCNN model training for three categories studied in this work. The classification between dark count and photon is demonstrated first, the following subsequent parts give the results of photon features (wavelength/polarization) recognition.
With the customized FCNN model ,figure \ref{fig:data_class} plots the comparison between original data and the predicted data of classifying dark count from normal photon. Here, the FCNN model consists of three layers, with 128, 64, and 32 neurons for each layer. As mentioned in the previous section, we use the Softmax method as the classifier, defining photon as 1 and dark count as 0 in the output of the neural network. It is clearly observed that predicted data and testing data match for the photon and dark counts with prediction accuracy can go up 100\%. Then the customized FCNN model is sufficiently used for the integration of FPGA in the real-time dark counts elimination of erbium photon emitter prototype test in the next sections.

\subsection{Recognition on Photon Wavelength and Polarization }
With the same methodology implemented for distinguishing dark count from the photon, the performance of recognizing different photon wavelengths and different polarization is demonstrated in this subsection, respectively. 

\begin{figure}[t]
    \centering
    \includegraphics[width=0.8\linewidth]{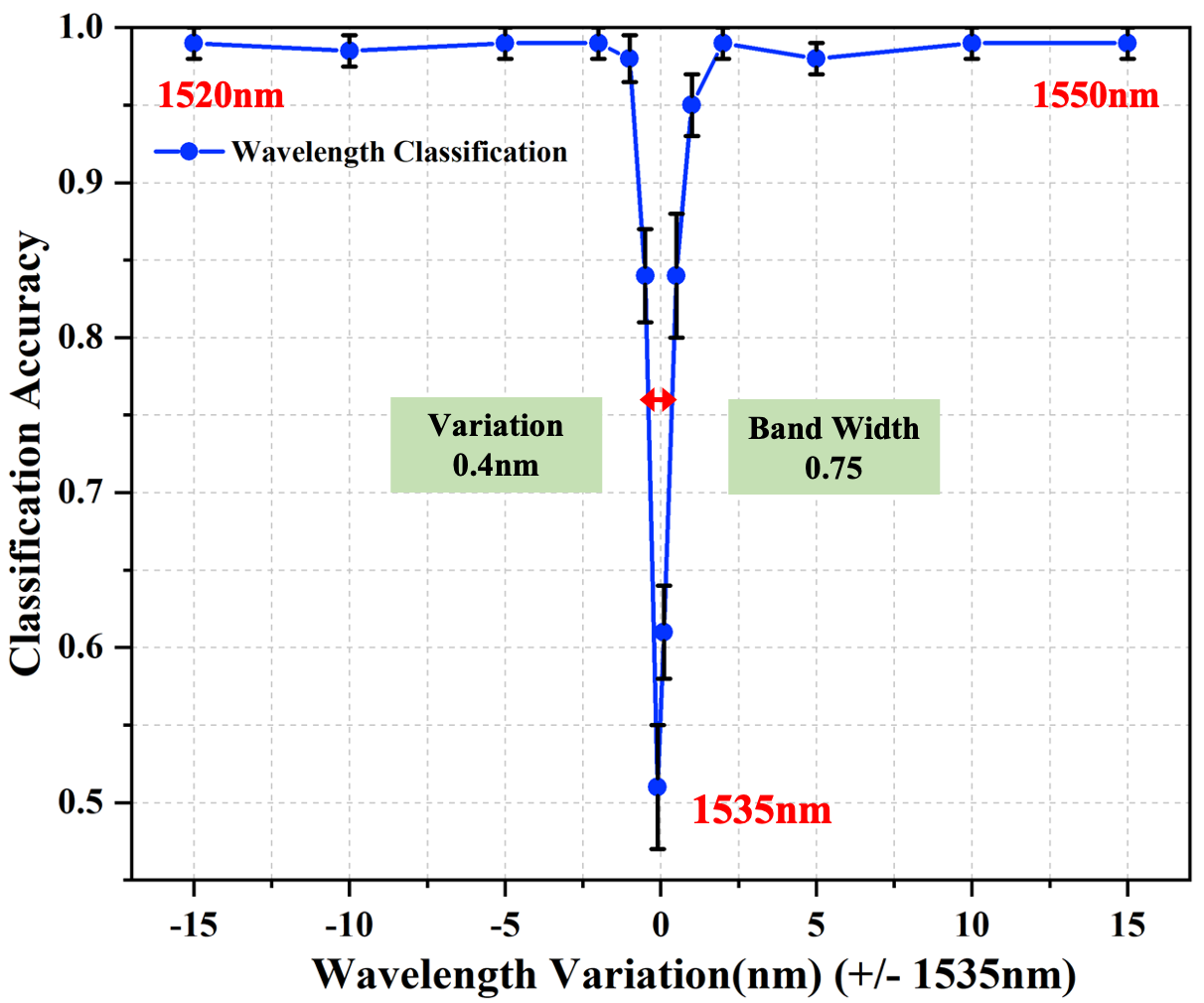}
    \caption{Test Accuracy of FCNN for Wavelength Classification Evaluation}
    \label{fig:photon_wavelength}
\end{figure}

\noindent{\bf Photon Wavelength Classification:} For the photon wavelength recognition evaluation, the laser source is tuned to obtain the data input for the model training with different wavelengths. The wavelength ranges from 1520nm to 1550nm, centered at 1535nm. Therefore, the performance of wavelength recognition between 1535 nm and other individual wavelengths in the range is evaluated. The evaluation results is plotted in Figure\ref{fig:photon_wavelength}. As demonstrated in the figure, the proposed \name system can recognize different of two wavelengths with a variation of 1 nanometer with accuracy up to 95\%. In addition, an accuracy of more than 75\% is obtained with a wavelength variation of 0.4 nanometers. In the end, it can be observed that with the wavelength variation increased, it is easier to recognize the difference, which obeys the physical rule that a wider wavelength introduces more difference in terms of photon energy.       

\noindent{\bf Photon Polarization Recongnization:} To demonstrate the generalizability of polarization recognition, the test accuracy between vertical polarization and horizontal polarization for multiple photon wavelengths.
Different from the accuracy analysis of different wavelength, we analyze the feature maps learned
by the FCNN using t-SNE visualization plots.

\begin{figure}[h]
    \centering
    \includegraphics[width=0.8\linewidth]{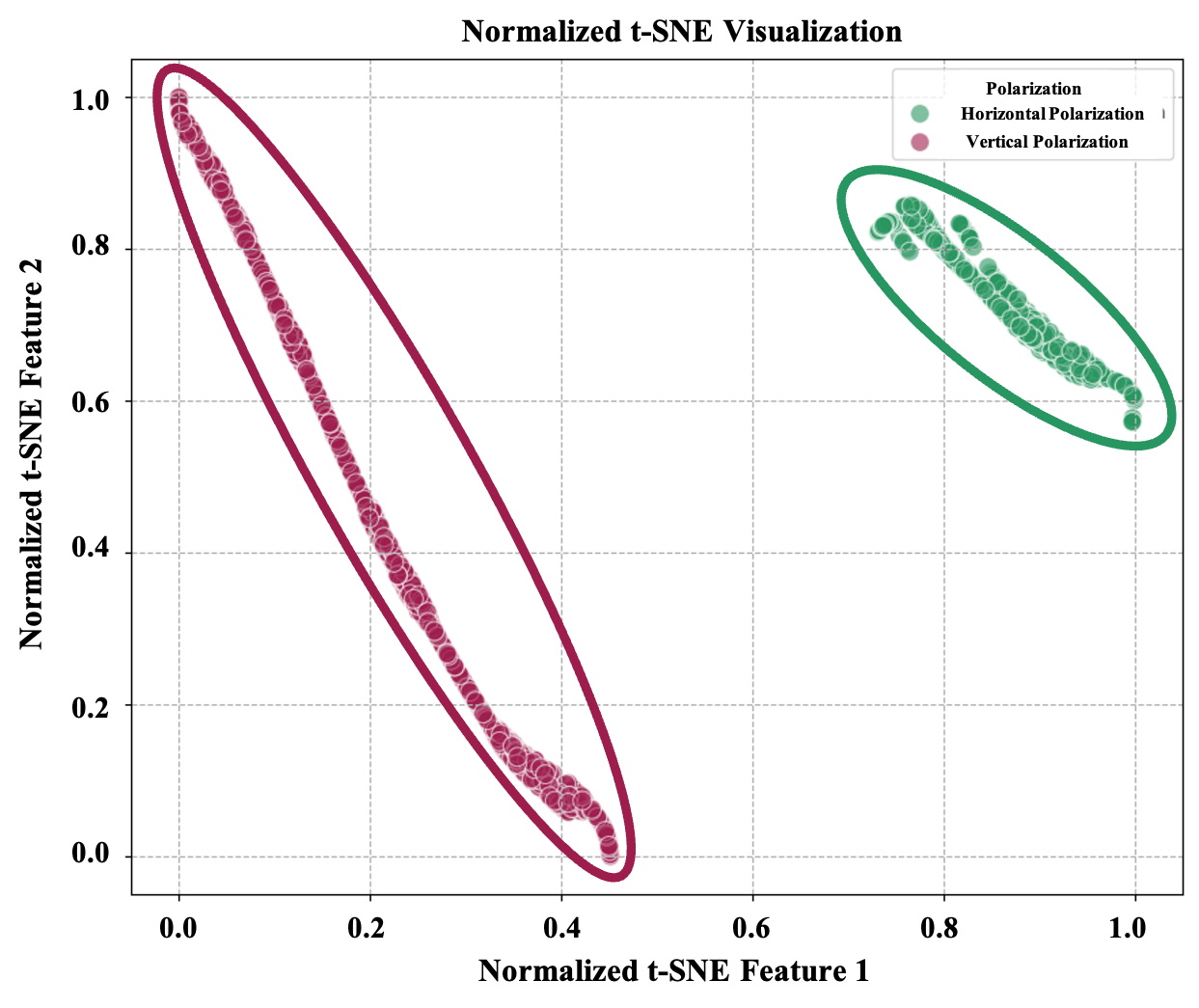}
    \caption{2D feature visualization with t-SNE plots of polarization classification.  We observe clear feature demarcation for two classes in the 2D feature visualization with t-SNE plots}
    \label{fig:photon_polarization}
\end{figure}

Figure\ref{fig:photon_polarization} plot t-SNE representation of FCNN when it is tested with vertical and horizontal polarization. In the  plot, we observe clear feature demarcation for these two classes in the 2D feature visualization with t-SNE. This indicates that the proposed FCNN integrated with the \name is able to learn the difference between polarization. Statically, the tested accuracy obtained in this work to recognize vertical polarization or vertical polarization is up to 100\%. 


\section{Evaluation of the Erbium Photon Emitter Prototype} This section demonstrates the initial results of the first prototype that integrates \name ~system into the erbium-based photon emitter. First, the erbium photon emitter architecture is shown. Then, the branching ratio and photolumincence performance are compared with dark count elimination and without elimination. 

\noindent
\subsection{Erbium Photon Emitter Architecture} 
\begin{figure}[h] 
\centering
\includegraphics[width=0.8\linewidth]{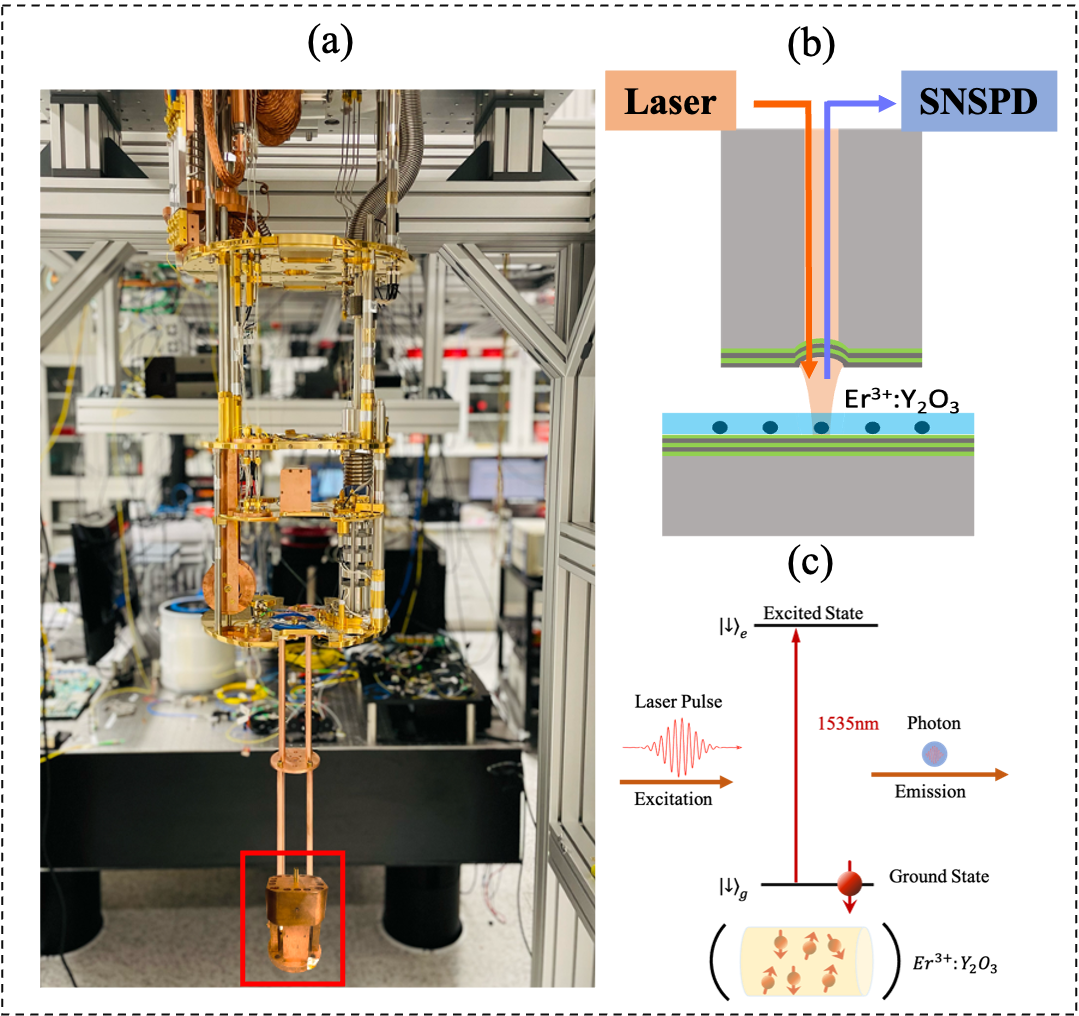}
\caption{Erbium single photon emitter system. a) operation environment with a temperature of 8.7 mK in dilute refrigerator; b) the conceptual architecture design with Fiber Coupled Fabry-Perot Cavity and erbium ion dopped in $Y_{2}O_{3}$ film; c) erbium ion energy diagram from the ground state to excited state.}
\label{fig:emitter_dilute}
\end{figure}

An erbium photon emitter prototype is developed for the systematic evaluation of \name system. In this work, this prototype aims to evaluate the function of dark counts, distinguishing, eliminating, and integrating laser beam control from the perspective of engineering techniques. The physical properties related to the erbium emitter will not be covered in detail in this work. This section will briefly illustrate the generic architecture design of the erbium photon emitter prototype. Similar to SNSPD, the erbium photon emitter operates at a dilution refrigerator at a temperature of 8.7 mK, illustrated as the red block of in \ref{fig:emitter_dilute}(a). Figure \ref{fig:emitter_dilute}(b) and (c) depicted conceptual architecture design and erbium ion energy diagram. Briefly, this emitter consists of two parts: 1) Fiber Fabry-Perot cavity (FFPC) to increase the photon emission rate\cite{ahmad2022tunable}, 2) erbium ion dopped in $Y_{2}O_{3}$ thin-films\cite{gupta2023robust}. 
In principle, the laser beam excites the erbium ion excited state from the ground state, and one photon is emitted while the erbium ion state goes back to the ground state, where the emitted photon will be detected by SNSPD and recorded by FPGA. The \name in the prototype test provides real-time whole system control, which includes the laser beam control, detection time control with the consideration of emission time, photon/dark count classification, and dark count elimination. Along with the erbium photon emission prototype, the laser locking method based on Pound-Drever-Hall (PDH)\cite{black2001introduction} was implemented stabilize the laser frequency generated by TOPTIC laser source, and then increase the photon emission efficiency.

\subsection{Erbium Photoluminescence Spectroscopy}
\begin{figure}[h]
    \centering
    \subfloat[]{
    \includegraphics[width=0.9\linewidth]{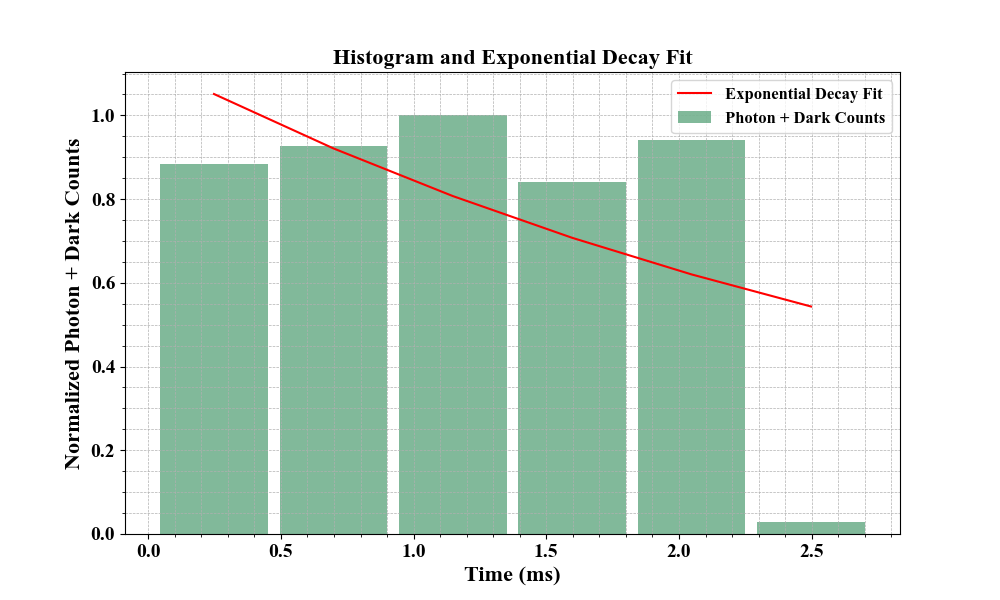}
    \label{subfig:photon_dark}
    }
    \vspace{-0.2cm}
    \subfloat[]{
    \includegraphics[width=0.8\linewidth]{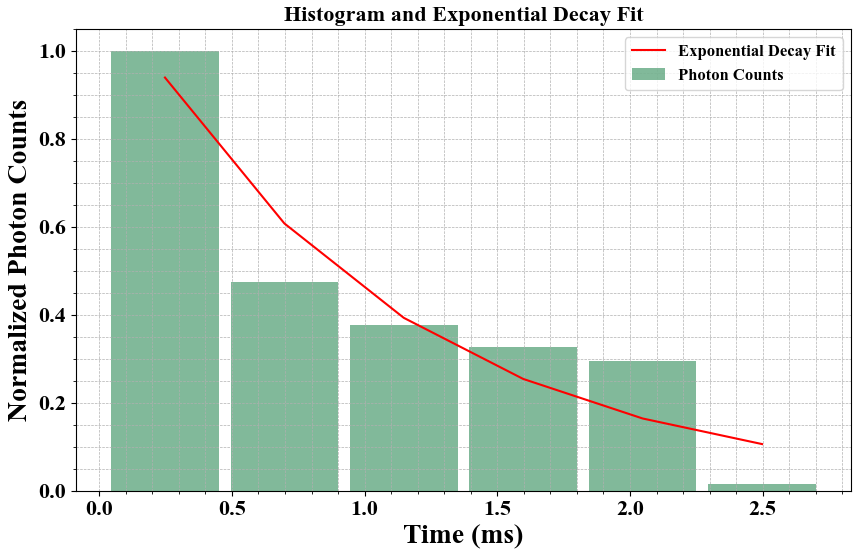}
    \label{subfig:photon}
    }
    \caption{Comparison of photoluminescence decay with dark count and dark count elimination }
    \label{fig:his_photon_dark}   
\end{figure}

This section demonstrates the evaluation of photoluminescence performance improvement due to the dark elimination with the proposed \name system. In the prototype test, the developed neural network model uploaded on the FPGA conducted the real-time processing of each detection. Meanwhile, the FPGA is utilized for control the whole system via the laser pulse control. Figure\ref{fig:his_photon_dark} plot photolumincence comparison between dark count elimination and dark count exists. From the results, it can be observed that a clear exponential decay trend of emitted photons with fewer errors was obtained after eliminating the dark count through \name. 
Statistically, we calculate the RMS error for both scenarios with the fitted exponential curve as the reference, and the results show that there is a 2.9 time improvement via eliminating the dark count with the proposed \name.     

\section{Conclusion}
This work proposes \name, a smart quantum detection system that utilizes a neural network model to classify and recognize multi-scale physical information of individual detection. We demonstrate complete system design, theoretical analysis, experimental performance evaluation, and practical prototype testing. The experimental evaluation shows that the proposed developed neural network model of \name achieves up to 100\% accuracy for the recognition of different photon wavelength and polarization, and for eliminating dark counts. The practical prototype testing with erbium photon emitter demonstrates more than 2.9 times improvement associated with rms errors analysis for the photoluminescence measurement, which verifies the capability of \name system for dark count elimination. The proposed \name maintains the integrity of the current quantum network and is ready to integrate seamlessly with today's quantum network in the future. 

\section{Acknowledgements}
\label{sec:ACK}
This work is supported PHOTONIDs, Inc. and U.S. DOE Office of Science-Basic Energy Sciences, under Contract No. DEAC02-06CH11357.


\end{document}